\documentclass[twocolumn]{aastex631}
\usepackage{amsmath}
\usepackage{natbib}

\shorttitle{QQO on Jupiter}
\shortauthors{Lian et al.}

\graphicspath{{./}{figures/}}

\begin{document}

\title{Jupiter's equatorial quasi-quadrennial oscillation forced by internal thermal forcing}

\correspondingauthor{Yongyun Hu}
\email{yyhu@pku.edu.cn}

\author{Yuchen Lian}
\affiliation{Laboratory for Climate and Ocean-Atmosphere Studies, Department  of  Atmospheric  and  Oceanic  Sciences,  School of Physics, Peking University, Beijing 100871, People’s Republic of China}
\affiliation{Key Laboratory of Planetary Sciences, Shanghai Astronomical Observatory, Chinese Academy of Sciences, Shanghai 200030, People’s Republic of China}

\author{Xianyu Tan}
\affiliation{Tsung-Dao Lee Institute, Shanghai Jiao Tong University, 520 Shengrong Road, Shanghai 201210, People’s Republic of China}
\affiliation{School of Physics and Astronomy, Shanghai Jiao Tong University, 800 Dongchuan Road, Shanghai 200240, People’s Republic of China}

\author{Yongyun Hu}
\affiliation{Laboratory for Climate and Ocean-Atmosphere Studies, Department  of  Atmospheric  and  Oceanic  Sciences,  School of Physics, Peking University, Beijing 100871, People’s Republic of China}

\begin{abstract}
Observations have shown that there exists downward propagation of alternating westward/eastward jets in Jupiter's equatorial stratosphere, with a quasi-period between four and six years. This phenomenon is generally called the quasi-quadrennial oscillation (QQO). Here, we simulate the QQO by injecting isotropic small-scale thermal disturbances into a three-dimensional general circulation model of Jupiter. It is found that the internal thermal disturbance is able to excite a wealth of waves that generate the equatorial QQO and multiple jet streams at middle and high latitudes of both hemispheres. The dominant wave mode in generating the QQO-like oscillation is that with a zonal wavenumber of 10.  Inhomogeneous evolution of potential vorticity favors the emergence of the off-equatorial zonal jets. The off-equatorial jets migrate to the equator, strengthen the deep equatorial jets, and result in the prolonging of the QQO-like oscillations.
\end{abstract}


\section{Introduction} 
\label{sec:intro}

Infrared imaging observations showed that there existed periodic changes of temperature anomalies in Jupiter's stratosphere in the past decades \citep[e.g.][]{orton-etal-1991,friedson-1999,flasar-etal-2004,simon-etal-2006,giles-etal-2020,antunano-etal-2021}. The equatorial temperature anomalies alternate between cold and warm states of ${\rm \pm 5}$ ${\rm K}$, with a quasi-period between 4 and 6 years. Associated with the temperature anomalies are vertically stacked eastward and westward zonal jets, with velocities variations of about ${\rm \pm 100}$ ${\rm m}$ ${\rm s^{-1}}$ \citep{orton-etal-1991,orton-etal-1994}. This phenomenon is termed the quasi-quadrennial oscillation (QQO) \citep{leovy-etal-1991}. Additionally, the stratospheric ethane's asymmetry with longitude and the equator may indicate the presence of QQO's circulation cells \citep{fletcher-etal-2016,fletcher-etal-2017}.

\citet{leovy-etal-1991} pointed out that the QQO closely resembles the Earth's quasi-biennial oscillation (QBO), in which equatorial zonal winds in the lower stratosphere oscillates between westward and eastward with a quasi-period of about $28$ months \citep{baldwin-etal-2001}. Theoretical and numerical studies demonstrated that the QBO is caused by the selective absorption of tropical waves \citep{lindzen-etal-1968,lindzen-1970,lindzen-1971,lindzen-1972}. It was shown that eastward (westward) equatorial gravity waves (both Kelvin and inertia-gravity waves) tend to be dissipated and absorbed when they encounter their critical layers, and that the deposited eddy momentum fluxes would accelerate the jets below the critical layers, while the westward (eastward) waves can propagate transparently in the mean flow. \citet{dunkerton-1985,dunkerton-1997} used a two-dimensional model in latitude and pressure and showed that the QBO can be driven both by equatorial trapped waves and small-scale gravity waves. It implies that temperature fluctuations in Jupiter’s equatorial stratosphere could be caused by a spectrum of breaking gravity waves \citep{young-etal-2005}.  

Evidence shows vigorous wave activities {in Jupiter's atmosphere}, especially in the equatorial region \citep[e.g.][]{rogers-mettig-2008, asay-davis-etal-2011,simon-etal-2012}. Most of these wave features appear within $\pm 5^{\circ}$ of latitude \citep{simon-etal-2015,orton-etal-2020}, and they exhibit diversity and complexity with various zonal wavenumbers, including waves with  {wavenumber between 17 and 70} \citep{harrington-etal-1996,cosentino-etal-2017b}. \citet{simon-etal-2012,simon-etal-2015} showed that these waves consist of wavenumbers $\ge 75$ and and have a phase speed of $101 \pm 3$ ${\rm m}$ ${\rm s^{-1}}$, which are likely inertia-gravity waves. The thermal wave patterns with wavenumbers $2$ to $15$, observed by \textit{Voyager}, have been identified as Rossby waves \citep{deming-etal-1997}. Rossby wave activities may generate dry downdrafts, participating in the formation of the regular arrays of cloud plumes \citep{friedson-2005,garcia-etal-2011}. The wave patterns also show an organization of zonal wavenumber $11$ to $13$ \citep{allison-1990}. The properties of waves in Jupiter's atmosphere have been observed from \textit{Galileo} and \textit{New Horizons}, with a vertical wavelength of $\sim 50$ ${\rm km}$ \citep{allison-atkinson-2001,arregi-etal-2009}, horizontal wavelengths of $\sim 10^3$ ${\rm km}$ \citep{reuter-etal-2007} and phase speed of $\sim 200$ ${\rm m}$ ${\rm s^{-1}}$ \citep{allison-atkinson-2001}. 

Sub-grid wave parameterization is widely utilized to simulate the QQO in large-scale models. \citet{friedson-1999} assumed that the equatorial waves are a combination of vertically propagating waves with eastward and westward velocities relative to the mean flow. They invoked a wave parameterization by choosing a flat gravity wave spectrum limited by maximum and minimum phase speeds to generate the momentum fluxes over waveguides. \citet{li-read-2000} followed \citet{friedson-1999} and their models established a QQO pattern with a period of $\sim$ 45 months. 
\citet{cosentino-etal-2017,cosentino-etal-2020} argued that the observed QQO signal extends up to 0.1 mbars which is lower than the top pressure in \citet{friedson-1999}. They utilized a stochastic gravity drag parameterization and reproduced the QQO thermal structures, which are similar to the structures observed by \textit{Cassini}. 

However, some aspects of Jupiter's QQO have not been explored. Many previous studies used two-dimensional models. These models lack 3D wave-flow interactions, in which wave generation and propagation and their interactions with the mean flow are parameterized and usually tuned to match certain aspects of observations. In contrast, 3D models have self-consistent wave-mean-flow interactions and are an alternative tool to elucidate the nature of the QQO. Secondly,  several interruptions of Jupiter's QQO have been observed. For example, the period of QQO is 5.7 years from 1980 to 1990, and the period of QQO is 3.9 years from 1996 to 2006 \citep{antunano-etal-2021}. In 2017,  a lag occurred with the downward signal of QQO,  delaying the phase of the QQO about a year \citep{giles-etal-2020}. These changes may be related to meteorological phenomena in the deep troposphere between 0.5 bar and 4 bar \citep{anderson-etal-2018}. However, the previous studies failed to simulate the variation of the QQO's period. We hope to provide some explanations for the variation of the QQO's period.

Here, we study Jupiter's QQO using a global general circulation model. Atmospheric waves are generated by  {isotropic} internal thermal perturbations near the radiative-convective boundary layers, instead of being parameterized. The wave-mean-flow interactions are self-consistently solved in our 3D model. The paper is organized as follows. The 3D model is described in Section \ref{sec:model}. In Section \ref{sec:results}, we present the simulation results and address the associated mechanisms. Conclusions and discussion are in Section \ref{sec:conclusions}.

\section{Model}
\label{sec:model}

\begin{figure}
\centering
\includegraphics[width=0.4\textwidth]{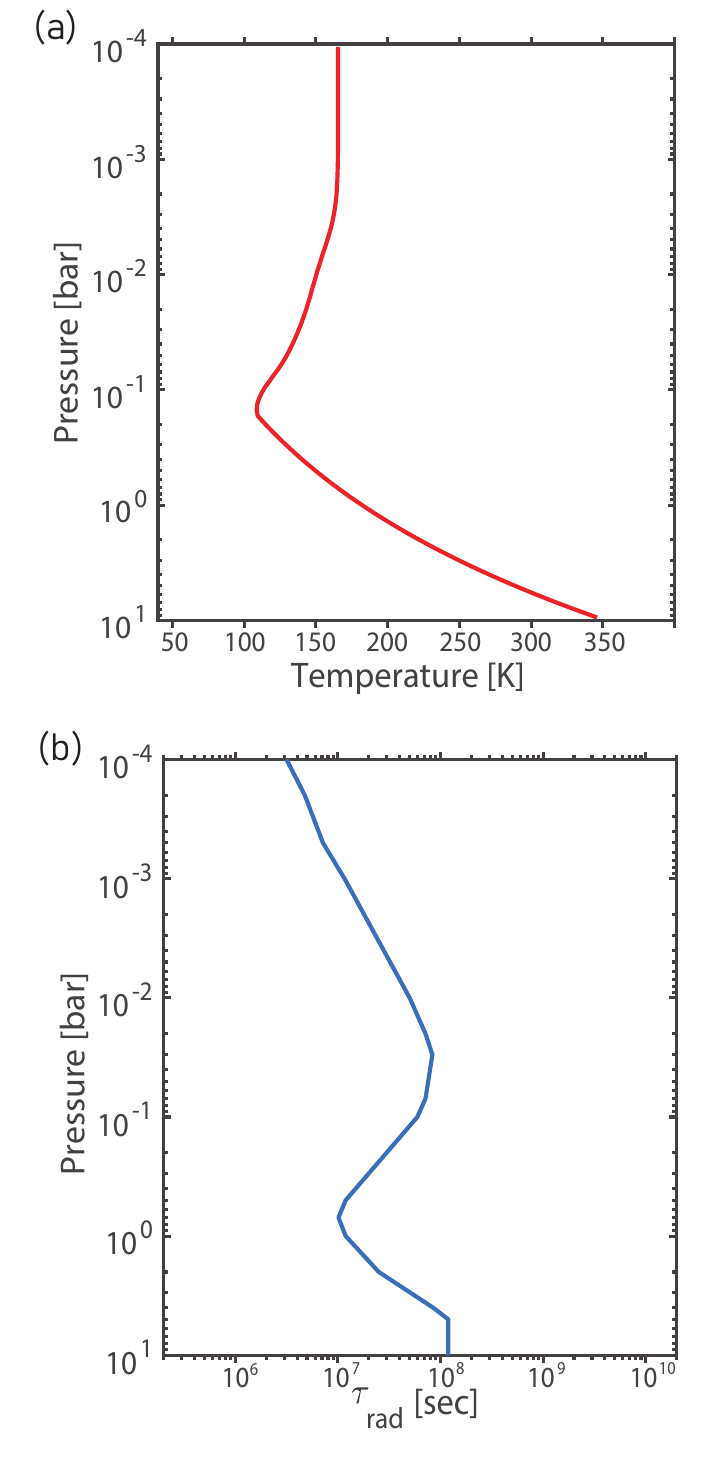}
\caption{Reference temperature (a) and radiative timescale profiles (b) used in our simulations.}
\label{fig:tp}
\end{figure}

The model used here is the MITgcm \citep{adcroft-etal-2004}. A Newtonian cooling scheme is applied to the thermodynamics equation, which relaxes the temperature to a reference temperature $T_{\rm ref}$ over a radiative timescale $\tau_{\rm rad}$. Newtonian cooling scheme has been widely used in previous studies for Jupiter \citep[e.g.][]{lian-showman-2010} and extrasolar giant planets \citep[e.g.][]{liu-showman-2013,mayne-etal-2014}. Both reference temperature $T_{\rm ref}$ and radiative timescale $\tau_{\rm rad}$ profiles are from  \citet{li-etal-2018}. The vertical dimension of the model extends from 10 bars at the bottom to $10^{-4}$ bars at the top. The profiles in \citet{li-etal-2018} are extended to 10 bars, by assuming a dry adiabatic temperature profile and a constant radiative timescale of ${\rm 2 \times 10^8}$ ${\rm s}$ at the bottom (Figure \ref{fig:tp}). The heating terms in the thermodynamic equation are therefore written as:
\begin{equation}
\label{thermdy}
{\frac{q}{c_p}}=-{\frac{T(\lambda,\phi,p,t)-T_{\rm eq} {(\phi,p)}}{\rm \tau_{rad}}}+S(\lambda,\phi,p,t+\delta t)
\end{equation}
where $q$ is the specific heating rate (${\rm W}$ ${\rm kg^{-1}}$), $c_p$ is the specific heat, $\lambda$ is longitude, $\phi$ is latitude, $p$ is pressure, and $t$ is time. The first term on the right-hand side is the Newtonian cooling term that represents radiative damping. We define the equilibrium temperature $T_{\rm eq}=\theta_{eq}(p/p_0)^{\kappa}$ and the equilibrium potential temperature as
\begin{equation}
\theta_{eq}(\phi,p)=\theta_{\rm ref}(p)+\delta \theta(\phi,p)
\end{equation}
where $p_0 = 10$ bars is the standard reference pressure, $\kappa = R/c_p$, $\theta_{\rm ref}$ is the reference potential temperature, and ${\rm \delta \theta}$ represents the latitudinal difference in equilibrium potential temperature corresponding to the solar irradiation. In our simulations, the hot equator and cold poles forced by solar insolation are parameterized as ${\rm \delta\theta = \Delta\theta}(p) \cos\phi$. ${\rm \Delta \theta}$ increases from 0 K at 3 bars to 10 K at $10^{-4}$ bar, roughly consistent with observations \citep{simon-etal-2006}.

The heating/cooling rate $S(\lambda,\phi,p,t+\delta t)$ in Equation (\ref{thermdy}) represents the internal forcing due to interior convective perturbations, mimicking overshooting and mixing effects across the radiative-convective  {boundary} at the bottom of stratified layers. We assume that convective perturbations are spatially isotropic and random in time. This scheme is originally developed in \citet{showman-etal-2019} for brown dwarfs and giant planets. \citet{tan-2022} applied this scheme to updated brown dwarf models, and \citet{lian-etal-2022} applied it to close-in gas giant planets. Here we adopt the scheme to the Jovian atmosphere. 

\begin{figure}
    \centering
    \includegraphics[width=0.4\textwidth]{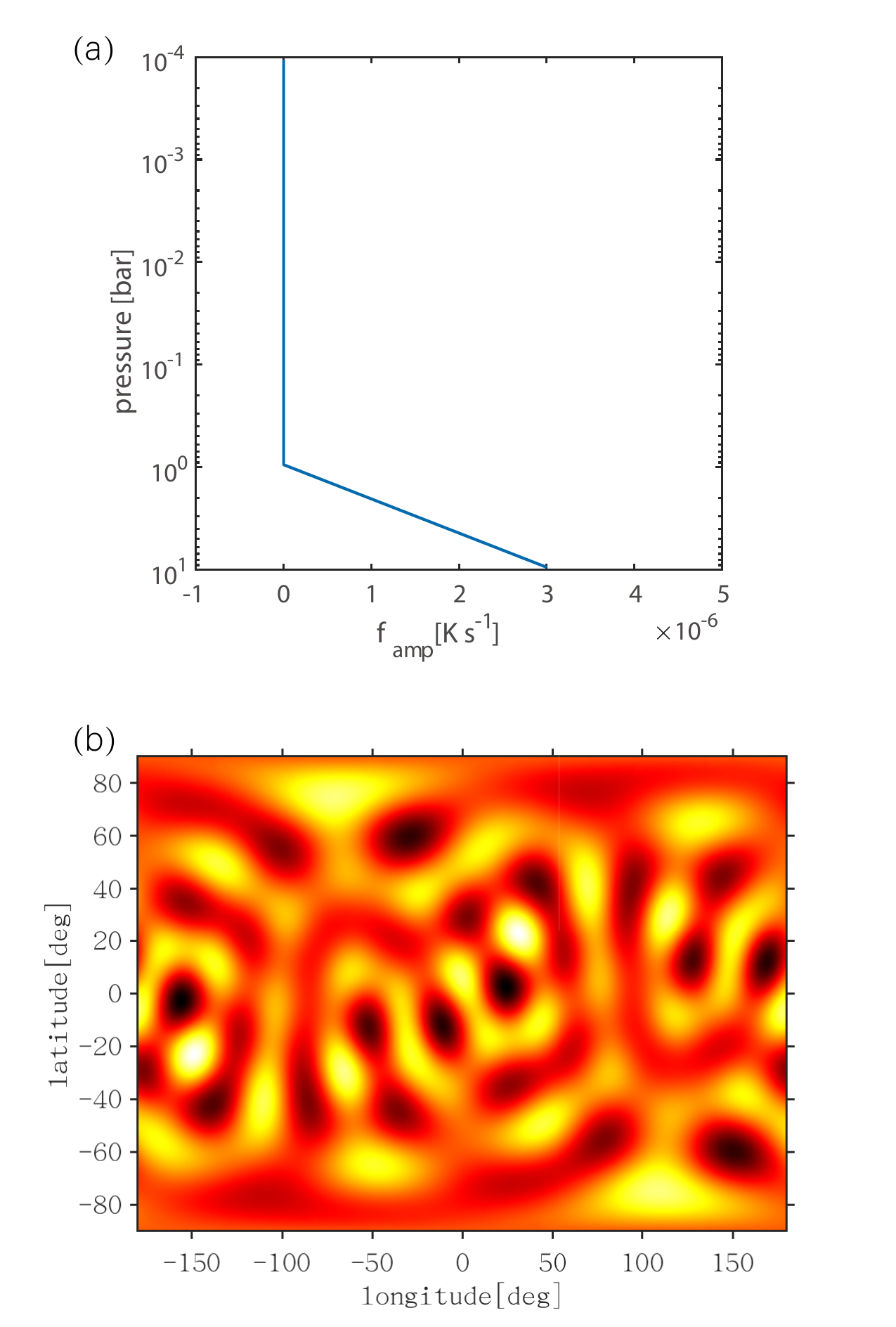}
    \caption{(a) The vertical structure of the thermal perturbation's amplitude $f_{\rm amp}$. (b) An example of the spatial pattern of $F(\lambda, \phi, p)$ in one layer shows the horizontal isotropic pattern. Here shows for $n_f=10$.}
    \label{fig:pattern}
\end{figure}

A spatial perturbation pattern $F$ is parameterized by globally isotropic heat sources at isobaric surfaces, which is numerically made by integrating all the spatial patterns of spherical harmonic functions with the wavenumber between 1 and $n_f$:
\begin{equation}
\label{forcbase}
F=f_{\rm amp}(p)\sum^{n_f}_{m=1}N^{m}_{n_f}(\sin\phi)\cos[m(\lambda+\psi_{m})]
\end{equation}
where $N^{m}_{n_f}(\sin\phi)$ is the normalized associated Legendre polynomials, $m$ and $\psi_{m}$ are the zonal wavenumber and a randomly chosen longitude phase, respectively. $f_{\rm amp}(p)$ is an internal forcing amplitude with unit of ${\rm K}$ ${\rm s^{-1}}$ with a vertical variation. $f_{\rm amp}(p)$ exponentially decreases with decreasing pressure and equals zero at the pressure $p_{\rm forctop} = 1$ bar. Figure \ref{fig:pattern} shows an example of spatial patterns of $f_{\rm amp}(p)$ and $F(\lambda, \phi, p)$. 

The heating/cooling rate $S$ evolves with time as a Markov process:
\begin{equation}
\label{markov}
S(\lambda,\phi,p,t+\delta t)=\sqrt{1-\alpha^2}S(\lambda,\phi,p,t)+\alpha F(\lambda,\phi,p)
\end{equation}
where ${\rm \alpha}$ is a de-correlation factor and equals ${\rm \delta t/\tau_{\rm s}}$, ${\rm \delta t}$ is the dynamical time step and $\tau_{\rm s}$ is the storm timescale of the internal forcing, representing the large-scale convective temporal organization. We are not able to accurately estimate the value of the storm timescale. However, observations provide us with some information. The thermal wave amplitude  {lasts} $10^5$ seconds on Jupiter \citep{deming-etal-1997}. Therefore, we set up the $\tau_{\rm s}$ as $10^5$ s in our canonical models and change the $\tau_{\rm s}$ from $10^6$ s to $10^3$ s in sensitivity tests.

We have to choose a forcing amplitude appropriate for Jupiter. The convective overshooting and mixing near the radiative-convective  {boundary} is expected to cause a temperature perturbation with an amplitude of $\Delta T_{\rm internal}$, 
We aim to generate the same amount of $\Delta T_{\rm internal}\sim 30$ K at 10 bars with theoretical $\tau_{\rm rad}\sim pc_p/4g\sigma T^3$ in our global-scale GCM by our forcing scheme, where $g$ is gravity and $\sigma$ is Stefan-Boltzmann constant. This can be estimated using the analysis in \citet{showman-etal-2019}, that $\Delta T_{\rm internal}$ is written as:
\begin{equation}
\frac{\Delta T_{\rm internal}}{\tau_{rad}}\sim f_{\rm amp}\sqrt{n_f}
\label{perturbation}
\end{equation}
$f_{\rm amp}$ and $\sqrt{n_f}$ should be restricted to obey the relationship in Equation \ref{perturbation} in order to keep temperature perturbations $\Delta T_{\rm internal}$ the same in different cases. $n_f$ is changed from $5$ to $40$ in different simulations to test the dominant wavenumber. In principle, we expect that the amplitude of internal forcing $f_{\rm amp}$ in our simulations is related to the real internal heat fluxes. Jupiter has $\sim$ 6 ${\rm W}$ ${\rm m^{-2}}$ internal heat flux \citep{ingersoll-1990,guillot-2005,li-etal-2018}, so we estimate that the thermal amplitude $n_f = 10$ yields $f_{\rm amp} \sim 4 \times 10^{-6}$ ${\rm K}$ ${\rm s^{-1}}$, and $n_f = 40$ yields $f_{\rm amp} \sim 2 \times 10^{-6}$ ${\rm K}$ ${\rm s^{-1}}$.

The internal forcing may be sufficiently strong to trigger super adiabatic regions, and we include a dry convective adjustment scheme to instantaneously remove the super adiabatic layers while conserving enthalpy. The stability criterion of a pair of adjacent stable layers is $T_{n+1} - T_k < C1_{n+1}(T_{n+1}+T_{n})$, where $n$ is the vertical index, $T$ is the temperature, and $C1_{n+1}=\kappa(p_{n+1}-p_n)/2p_{(n+1)/2}$. If any pair of adjacent layers are unstable, the temperature will be adjusted to be convectively neutral while conserving total enthalpy. Such process is repeated until the whole column satisfies the stability criteria \citep{manabe-strickler-1964}.

\begin{figure*}[!ht]
\centering
\includegraphics[width=0.65\textwidth]{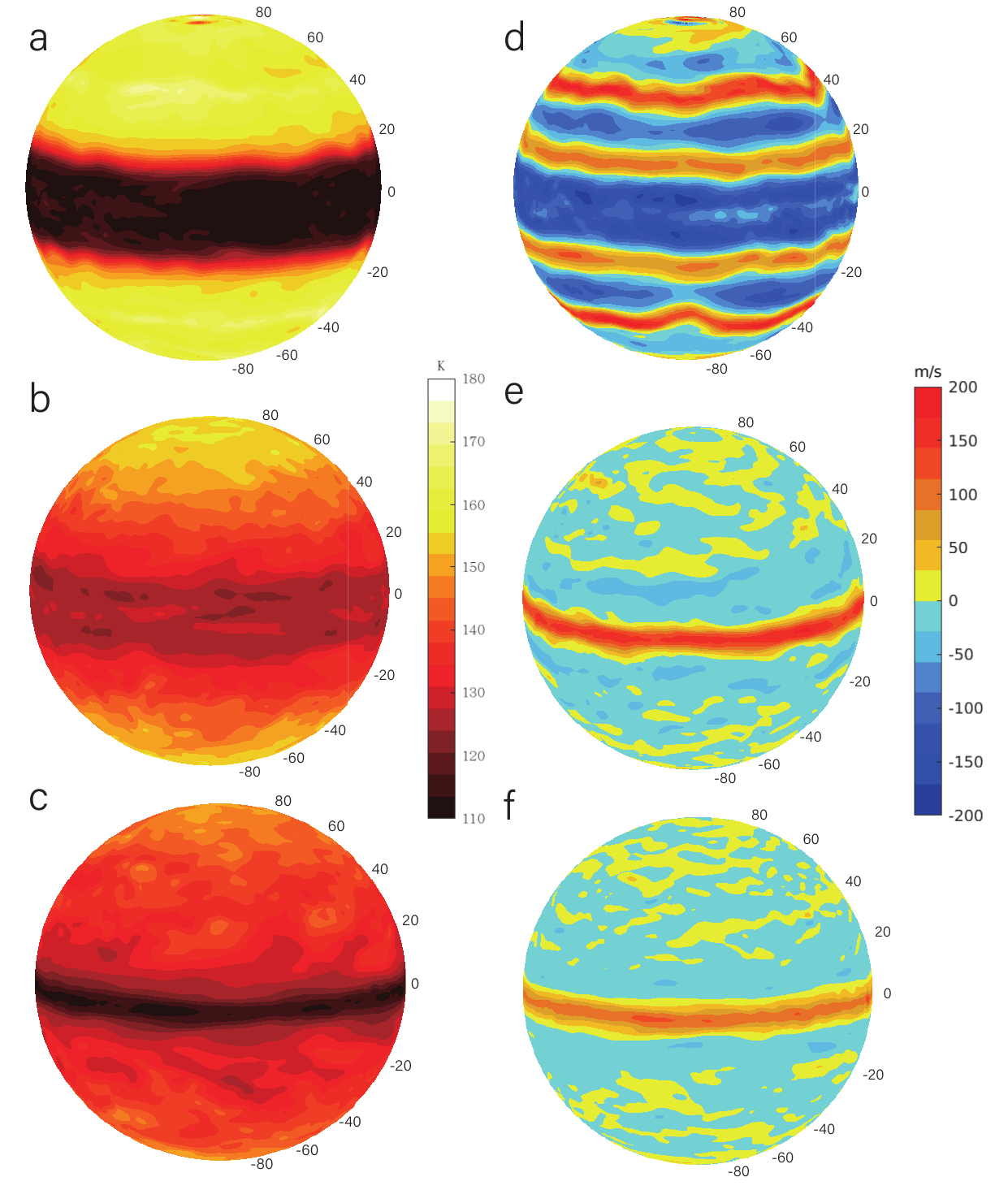}
\caption{Temperature (${\rm K}$, left panels) and zonal winds (${\rm m}$ ${\rm s^{-1}}$, right panels) at a pressure of 60 mbars in simulations with different bottom drag. These are snapshots at 10000 simulation days. The drag timescales are $10^7$ s (a), $10^6$ s (b), and $10^5$ s (c). The forcing amplitude is $4$ $\times$ $10^{-6}$ ${\rm K}$ ${\rm s^{-1}}$, storm timescale $\tau_{\rm s}$ is $10^5$ s, forcing wavenumber $n_f$ $=$ $10$, and other parameters are described in the text.}
\label{fig:sph-drag}
\end{figure*}

We apply a Reyleigh drag term $-\textbf{u}/\tau_{\rm drag}$ into the dynamic equations to mimic the effects of angular momentum mixing with the planetary interior where the magnetohydrodynamic drag is important \citep{liu-schneider-2010}, where $\textbf{u}$ is the horizontal velocity. The drag is linearly decreased with decreasing pressure. That is, $\tau_{\rm drag}$ increases from a certain timescale (e.g. $10^7$, $10^6$ s) at the bottom linearly with decreasing pressure to infinite at a certain pressure $p_{\rm drag,top} = 4$ bars, after which the atmosphere is drag-free at pressures $\leq p_{\rm drag,top}$. In order to diminish wave reflection at the upper boundary, we add a ``sponge'' layer for dissipation. This dissipation term is introduced as $-\nu\textbf{u}$, and $\nu$ is the same as the damping coefficient form in \citet{dowling-etal-1998}:
\begin{equation}
\nu(k)=\frac{1}{5\delta t}\frac{1}{2}(1-\cos(\pi\frac{k_{sp}+1-k}{k_{sp}})),
\end{equation}
where $k$ is the vertical index ($k=1$ at bottom) and $k_{\rm sp} = 10$ is the number of sponge layers.

The model has 100 vertical layers. The rotation period is set to be 9.84 hours, and planetary radius $R_J = 7.14 \times 10^{7}$ ${\rm m}$, $c_p=13000$ ${\rm J}$ ${\rm kg^{-1}}$ ${\rm K^{-1}}$ and $\kappa = R/c_p = 2/7$ appropriate to a hydrogen-dominated atmosphere. The gravity is set to 24.79 ${\rm m}$ ${\rm s^{-2}}$. We adopt a horizontal resolution of C128 in our cubed-sphere grid in MITgcm, corresponding to horizontal resolution 128 $\times$ 128 for each cube face ($0.7^\circ$ per grid in latitude and longitude). The timestep is 100 seconds for most cases. All simulations integrated until balance, approximately 7000-10000 Earth days.

\begin{figure*}[!ht]
\centering
\includegraphics[width=0.7\textwidth]{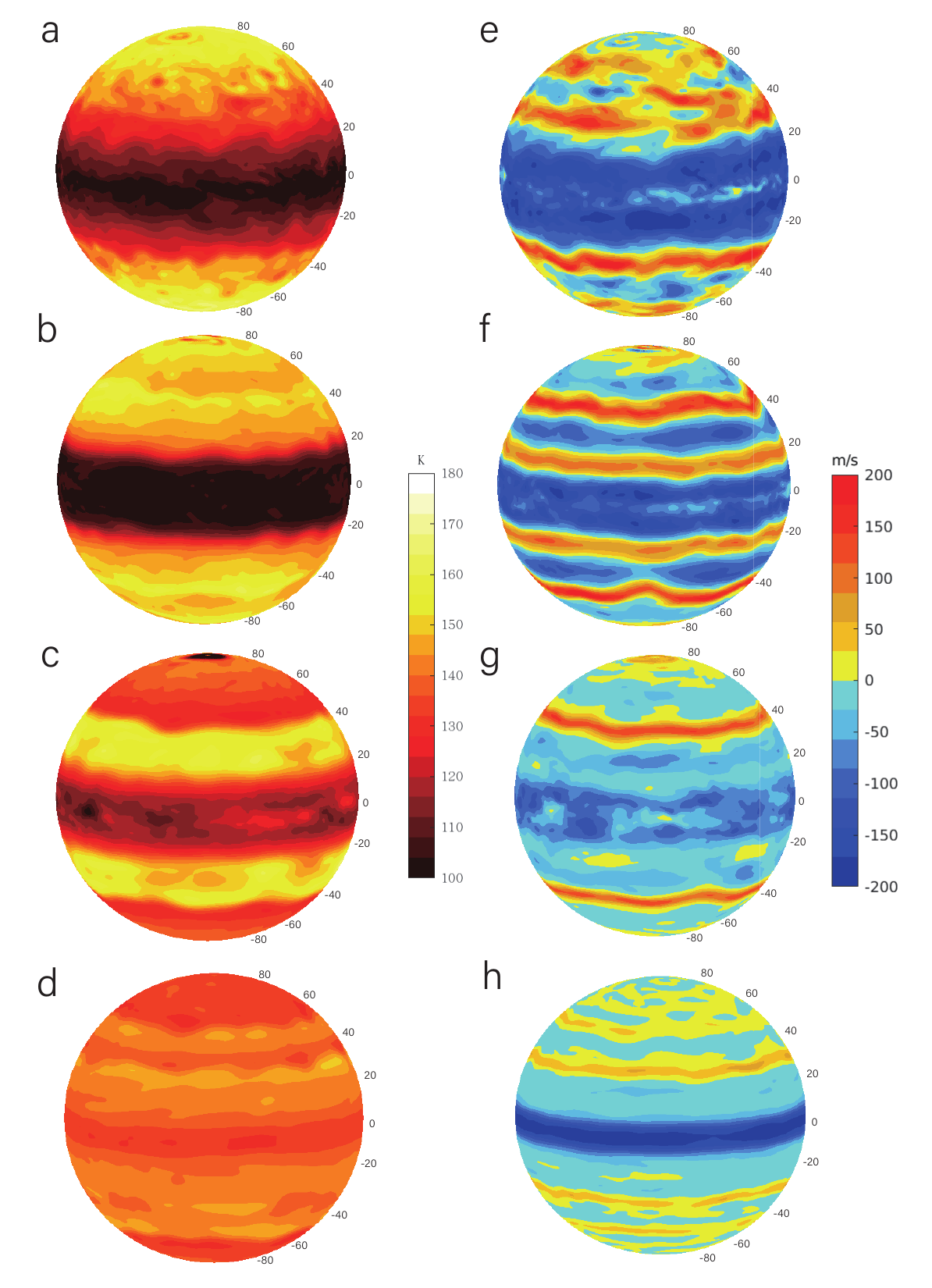}
\caption{Temperature (${\rm K}$, left panels) and zonal winds (${\rm m}$ ${\rm s^{-1}}$, right panels) at 60 mbars in four simulations with different $\tau_{\rm s}$. These are snapshots at 10000 simulation days. The four simulations are identical, except for different storm timescales: $10^6$ (a), $10^5$ (b), $10^4$ (c), and $10^3$ s (d). The forcing wavenumber $n_f$ is $10$ and the drag timescale $\rm \tau_{\rm drag} = 10^7$ s. Other parameters are the same as the parameters in Figure \ref{fig:sph-drag}.}
\label{fig:sph}
\end{figure*}

\section{Results}
\label{sec:results}

\subsection{Basic Flow Regime}
Figure \ref{fig:sph-drag} shows temperature and zonal wind patterns with different bottom drag timescales. The drag timescales decrease from $10^7$ s in Figure \ref{fig:sph-drag}a, d to $10^5$ s in Figure \ref{fig:sph-drag}c, f (the drag effects increase in strength from top to bottom). Figure \ref{fig:sph} shows temperature and zonal wind patterns with decreasing storm timescales from $10^6$ s in Figure \ref{fig:sph}a, e to $10^3$ s in Figure \ref{fig:sph}d, f. The key result is that the isotropic internal forcing results in zonal jets, and that the zonal jets are sensitive to drag and storm timescales. This is qualitatively in agreement with previous results in the context of isolated brown dwarfs and giant planets that lack equator-to-pole stellar irradiation difference \citep{showman-etal-2019,tan-2022}.

\begin{figure}
    \centering
    \includegraphics[width=0.3\textwidth]{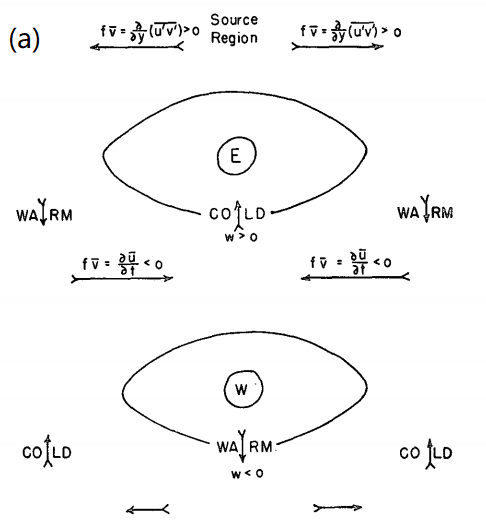}\\
    \includegraphics[width=0.4\textwidth]{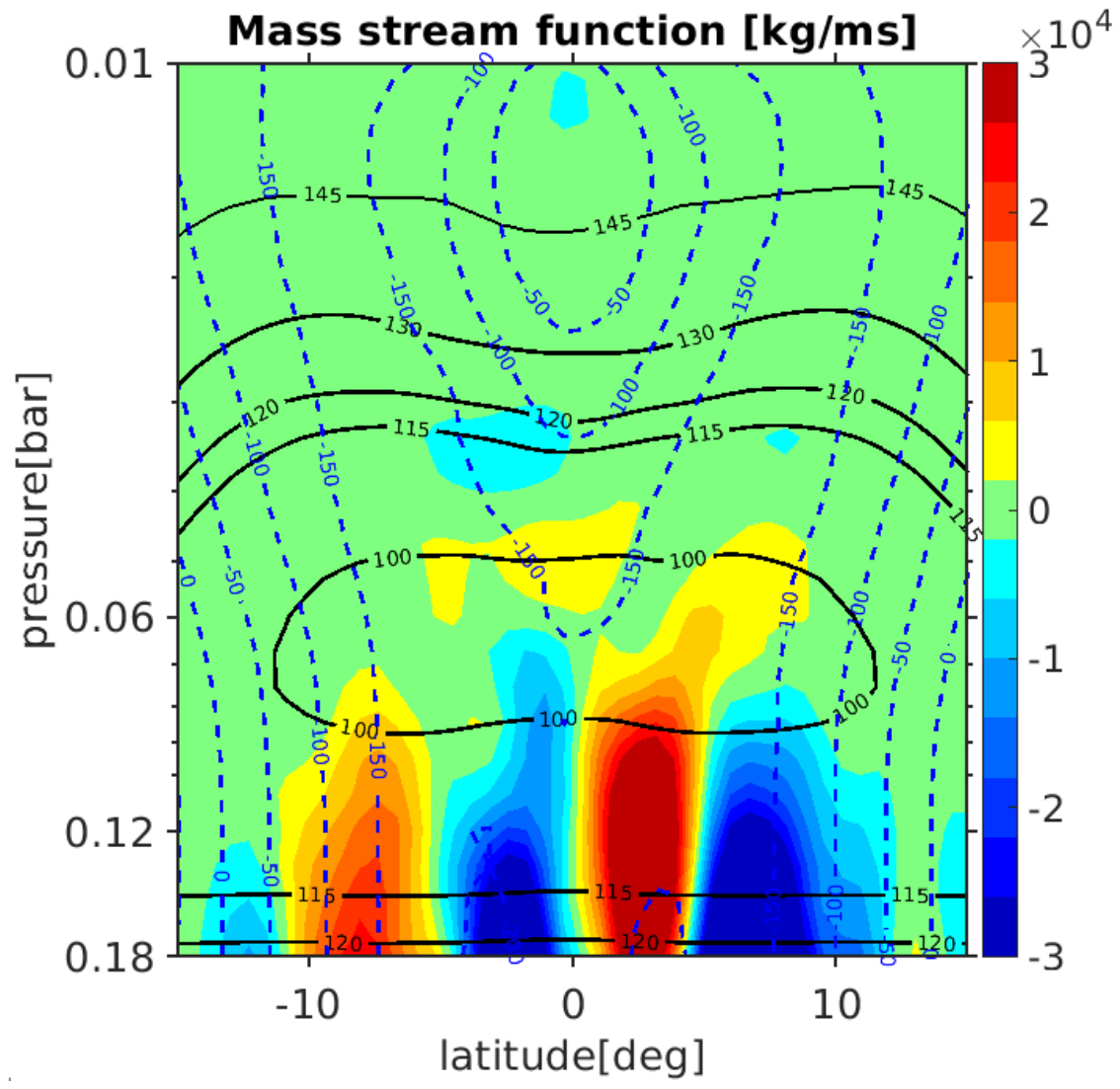}
    \caption{(a) The sketch depiction of the kinetic mechanism of equator-to-subtropical  temperature inverse caused by the QBO, from \citet{dickinson-1968}. E represents the eastward wind and W represents the westward wind. (b) The colormap shows the mass stream function at the equatorial region of cases with $\tau_{\rm s}=10^5$ and $\tau_{\rm drag}=10^7$ (same as Figure \ref{fig:sph}b). The black solid line shows the zonal mean temperature and the blue dash line shows the zonal-mean zonal wind.}
    \label{fig:inv}
\end{figure}

\begin{figure*}[!ht]
\centering
\includegraphics[width=0.8\textwidth]{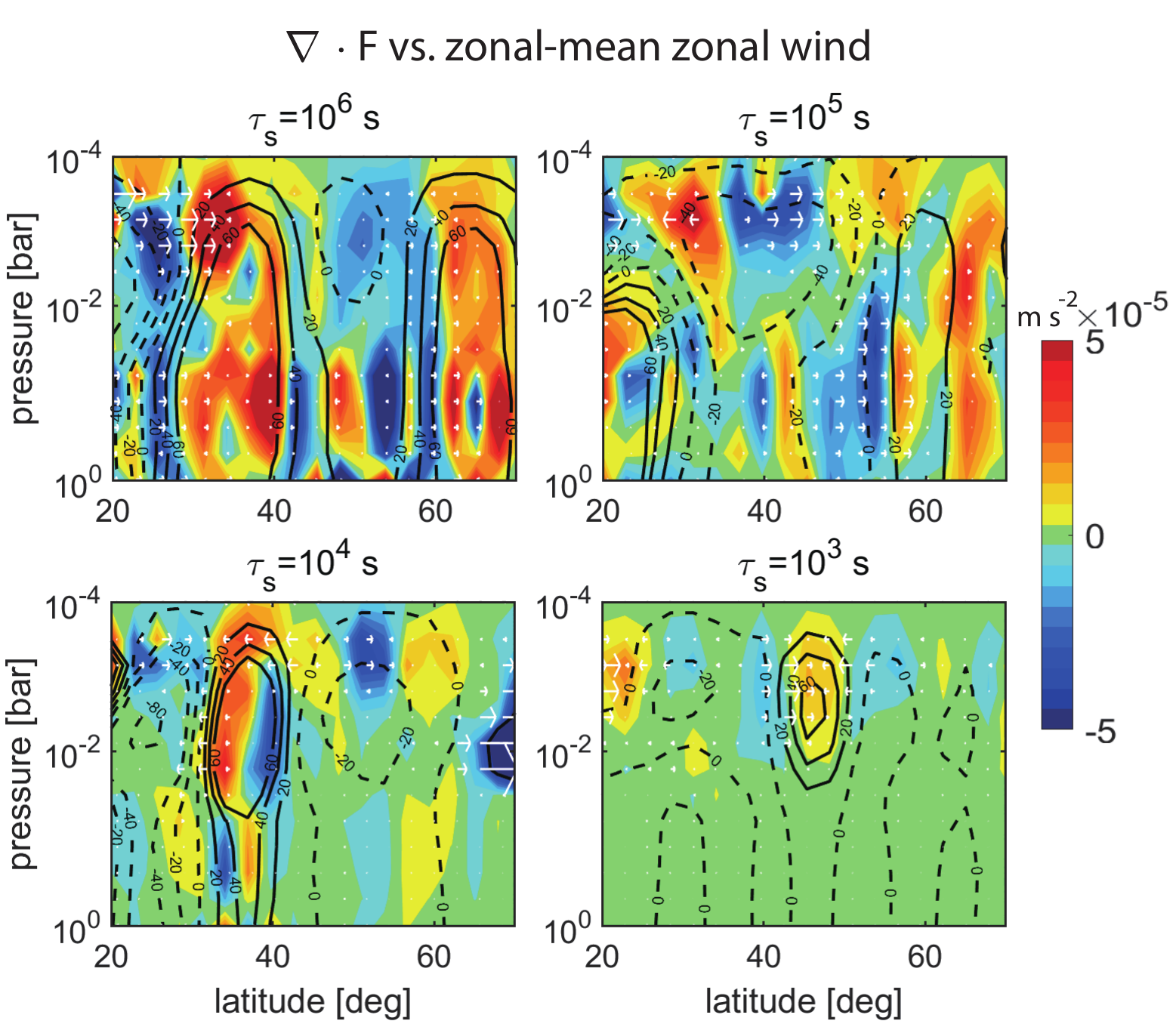}
\caption{Wave-induced acceleration (colormap) ($\nabla \cdot \textbf{F}$) at north hemisphere's mid- and high-latitudes with zonal-mean zonal wind (solid lines: positive and dash lines: negative) and Eliassen-Palm fluxes ($\textbf{F}$) (white arrows) of four $\tau_{\rm s}$ simulations. All quantities are on time averaging from 10000 to 10020 days. Meridional propagation of Rossby waves generates eastward and westward accelerations by eddies, leading to various zonal jets.}
\label{fig:meriwave}
\end{figure*}

The left panels of Figure \ref{fig:sph-drag} and \ref{fig:sph} show a temperature anomaly in the subtropics of opposite sign to that at the equator around 60 mbars, with a colder equator about $100$ ${\rm K}$ and warmer subtropics and poles $\sim 160$ ${\rm K}$, which is caused by overturning circulation that is generated by the equatorial wave accelerations. The same meridional temperature variation has been found in the QBO simulations on Earth \citep[e.g.][]{dickinson-1968,plumb-etal-1982b,baldwin-etal-2001}. 

If the heating is homogeneous around the latitude circle, the zonal-mean momentum equation is:
\begin{equation}
    \frac{\partial \bar{u}}{\partial t}-f\bar{v}=a_x
\label{rev}
\end{equation}
$a_x$ can be the acceleration momentum source, which comes from the eddy-momentum transports or the friction. Coriolis force is weak near the equator, so the second term is negligible, and all the accelerations work on the background flow, leading to the time evolution of the zonal jets. However, Coriolis force increases poleward, becoming the main part to balance the momentum acceleration. An overturning circulation is driven by the eddy accelerations, forcing the air to ascend under the eastward jets and descend above the westward jets. A colder equator appears in the  {cyclonic shear} zone due to the adiabatic cooling by ascending motions, and the warmer equator appears in the  {anti-cyclonic shear} zone. In brief, the upwelling with equatorial cold anomaly accompanies the eastward equatorial wind, while the downwelling with equatorial warm anomaly accompanies the westward equatorial wind \citep{dickinson-1968}. Figure \ref{fig:inv}a shows a schematic for the mechanism causing this meridional temperature variation. Figure \ref{fig:inv}b shows the matching of the eastward wind (the blue dash line circle), cold center (the black solid line circle), and the overturning circulation, which are shown by the stream function colormap. 
 
We have ruled out the Kelvin-Helmholtz instabilities of vertical wind shears to generate waves. We confirm that the Richardson number $R_i=N^2/(\partial u/\partial z)^2$ is far greater than $1$ in our simulations, where $N$ is Brunt-Vaisala Frequency.

Meridional propagations of Rossby waves help generate and spatially inhomogenized zonal jets by breaking the absolute vorticities into PV staircases \citep{dritschel-mcintyre-2008,dunkerton-etal-2008}. The absolute vorticity is the sum of the relative vorticity and the Coriolis parameter, providing an approximation of potential vorticity (PV). A particularly useful form of analyzing angular momentum transport to the mean flow by atmospheric waves was developed by \citet{eliassen-palm-1961} that involves analyzing the so-called Eliassen-Palm (EP) flux. The EP flux equation in the quasi-geostrophic balance reads \citep[e.g.][]{andrews-etal-1987}:

\begin{equation}
    \bar{\textbf{F}}=\bar{F}_y+\bar{F}_z=-\rho \overline{u'v'}\textbf{j}+\rho f_0\frac{\overline{v'\theta'}}{\bar{\theta}_z}\textbf{k}
\label{eqn:ep}
\end{equation}
primes denote the residuals from the zonal average (overbars). The acceleration of the mean flow is corresponding to the convergence of the EP flux $\nabla \cdot \bar{\textbf{F}}=\frac{\partial \bar{F}_y}{\partial y}+\frac{\partial \bar{F}_z}{\partial z}$. We also have $\overline{v'q'}=\nabla \cdot \bar{\textbf{F}}$, where $q$ is PV, $q'$ is the perturbations relative to the zonal-mean part $\bar{q}$ from total PV, and $\overline{v'q'}$ is the mean meridional flux of PV. The EP flux is corresponding to the meridional flux of PV.

For linearized planetary Rossby waves, the group velocities $\textbf{c}_g$ are related to the EP fluxes via \citep{vallis-2006}:
\begin{equation}
    \bar{\textbf{F}}=\textbf{c}_g \bar{A}
    \label{eprossby}
\end{equation}
The quantity $\bar{A}={p\bar{q'}^2}/{2\bar{q}_y}$ is the mean wave activity density. That is, the group velocities of planetary Rossby waves are parallel to the EP fluxes, and the meridional PV fluxes are corresponding to the meridional Rossby wave group velocities. The Rossby waves are more easily breaking in the regions with weak PV gradients. When the Rossby waves break, the EP fluxes transport momentum to the mean flow \citep{hoskins-etal-2014}. Rossby waves breaking causes PV mixing by the eddies and weaken the PV gradients further \citep{dritschel-mcintyre-2008,marcus-shetty-2010}. As a result, the positive feedback continues and the PV gradient moves to a state with a high-low contrast between  {latitudes, leading to the formation of zones and belts}. The sharp edges between the inhomogeneous PV are corresponding to the maximum speeds of the zonal jets.

We could obtain a general idea of the amplitude and propagating directions of Rossby waves from the white arrows of EP fluxes in Figure \ref{fig:meriwave}, that the arrows represent the group velocity directions and their lengths are proportional to the group velocities (Equation \ref{eprossby}). The shorter arrows are consistent with the center of zonal jets with weaker PV gradients, showing more wave breaking. When waves reach the critical latitudes where their phase speeds are equal to the mean speeds of jets, their meridional group velocities tend to zero, which prevents further propagation \citep{orourke-vallis-2016}. While the regions close to zero wind contours have larger EP fluxes, showing the existence of strong meridional Rossby wave propagations between the strips of PV. The convergences of EP fluxes are shown in colormap in Figure \ref{fig:meriwave}, indicating that the waves are absorbed. Wave absorption causes wave-induced acceleration of the same sign as the background flow, in which the red color shows eastward and the blue color shows westward. The PV is forced to organize staircase patterns, promoting the formations of the zonal jets. We have demonstrated that this mechanism also works in a full 3D model by analyzing the EP flux convergence and divergence.

After explaining the formations of the jets, we analyze the factors that affect the jet distributions. Firstly, the rotation rate affects the widths of the jets. Generally speaking, the widths of the jets are determined by the  {Rhines} scale $L_R = \sqrt{U/\beta}$, where $U$ is the  {eddy} wind speed and $\beta$ is the meridional gradient of the Coriolis parameter. When the motion scale reaches the  {Rhines} scale, the eddy fields will be affected by the gradients of the Coriolis force, and the $\beta-$effect induces the self-organization of the jets \citep{rhines-1975}. The jet widths decrease with the increase of the rotation rate. Second, Figures \ref{fig:sph-drag}d-f show that the maximum off-equatorial jet speeds decrease from about $200$ ${\rm m}$ ${\rm s^{-1}}$ with the decreasing ${\rm \tau_{drag}} = 10^7$ s to about $20$ ${\rm m}$ ${\rm s^{-1}}$ with ${\rm \tau_{drag}} = 10^5$ s, associated with increasing bottom drag. The off-equatorial jets are more sensitive to the bottom drag, and the zonal jets only develop at the equator with strong bottom drag. In geostrophic balance, horizontal divergence is larger at low latitude \citep[e.g.][]{showman-etal-2013}, that
\begin{equation}
    \nabla \cdot \textbf{u} = \frac{\beta}{f}v = \frac{v}{R_J\tan\phi}
\end{equation}
At low latitudes where $\tan \phi$ is small, larger divergence allows larger vertical movements, enhancing the horizontal temperature differences and generating waves more easily, promoting the jet formations at low latitudes. 

The equatorial jet speeds increase at first  {with $\tau_{\rm drag}$} and then decrease, even if they all are in the eastward phase. The maximum equatorial jet speeds increase from about $50$ ${\rm m}$ ${\rm s^{-1}}$ in weak drag case with $\tau_{\rm drag} =$ $10^7$ ${\rm s}$ of Figure \ref{fig:sph-drag}d to about $180$ ${\rm m}$ ${\rm s^{-1}}$ of Figure \ref{fig:sph-drag}e and then decrease to about $100$ ${\rm m}$ ${\rm s^{-1}}$ of Figure \ref{fig:sph-drag}f. The off-equatorial westward jets near $\pm 15^\circ$ are suppressed a lot, and the equatorward momentum exchange is also weakened, so the equatorial eastward jets increase from Figure \ref{fig:sph-drag}d to e; Then the drag increases to such an extent that the eastward equatorial jets are damped and thus weaken again from e to f.

In Figure \ref{fig:sph}, we keep the bottom drag and amplitude of perturbation the same, i.e. $\tau_{\rm drag} =$ $10^7$ ${\rm s}$ and $f_{\rm amp} =$ $4 \times 10^{-6}$ ${\rm K}$ ${\rm s^{-1}}$, but decrease the storm timescale $\tau_{\rm s}$ from ${\rm 10^6}$ s at the top Figure \ref{fig:sph}a to ${\rm 10^3}$ s at the bottom Figure \ref{fig:sph}d. The simulation with a long ${\rm \tau_{\rm s}} = 10^6$ s shows much more turbulence with eddies at mid-to-high latitudes while the short ${\rm \tau_{\rm s}}$ simulations only show robust jets. The peak negative wind speeds (westward jets) are ${\rm \sim -200}$ ${\rm m}$ ${\rm s^{-1}}$ at the equator of all the simulations, but the eastward off-equatorial jets drastically decrease speeds from about $200$ ${\rm m}$ ${\rm s^{-1}}$ to $100$ ${\rm m}$ ${\rm s^{-1}}$ with decreasing ${\rm \tau_{\rm s}}$. 

We suggest an explanation for the relationship between $\tau_{\rm s}$ and the off-equatorial jet structure: the larger $\tau_{\rm s}$, the stronger parameterized convective heating, which results in larger EP fluxes, leading a larger zonal acceleration maintaining the zonal jets. Detailed investigations are shown in Subsection \ref{time}.

To summarize, with the decreasing $\tau_{\rm drag}$ and increasing drag in Figure \ref{fig:sph-drag}, the jets storm timescale, showing that the drag removes the momentum out of zonal jets, confining the jets at low latitudes. With the decreasing $\tau_{\rm s}$ in Figure \ref{fig:sph} and Figure \ref{fig:meriwave}, the amplitude of wave-induced acceleration shows a declining tendency, consistent with the lower speeds of zonal jets.

\begin{figure*}
\centering
\includegraphics[width=0.9\textwidth]{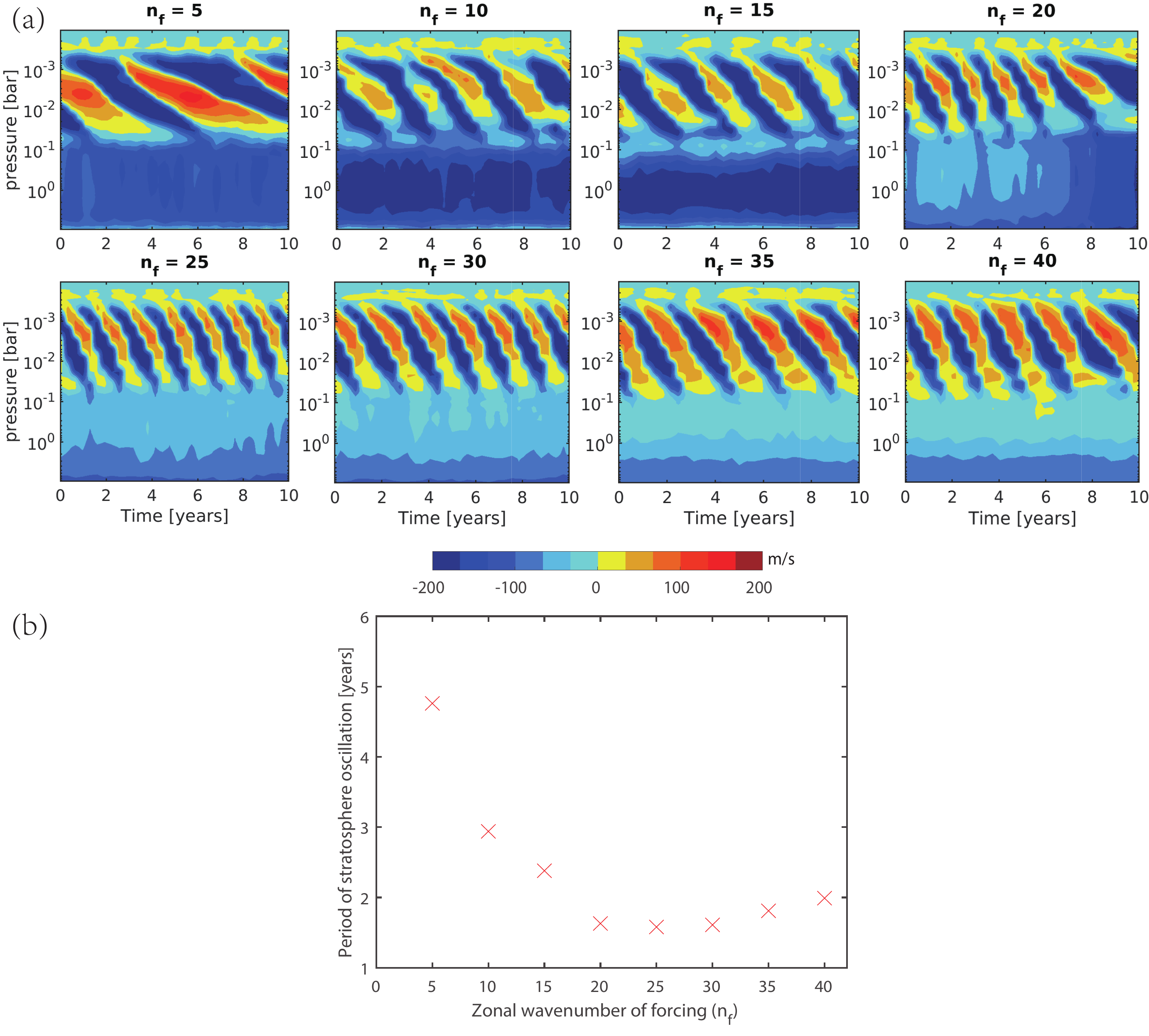}
\caption{(a) Zonal winds at the equator with time shown eastward (red region) and westward wind (blue region) modes between ${\rm \pm 200}$ ${\rm m}$ ${\rm s^{-1}}$. The titles $n_f$ show different zonal forcing wavenumbers. In all simulations, $f_{\rm amp}\sqrt{n_f}= 1.3 \times 10^{-5}$ ${\rm K}$ ${\rm s^{-1}}$, the storm timescale $\tau_{\rm s}=10^5$ s, the drag timescale $\tau_{\rm drag} = 10^7$ s, and other parameters are as described before. Color bar unit is ${\rm m}$ ${\rm s^{-1}}$. The period unit is Earth year. (b) A scattering of the relationship between the zonal forcing wavenumber and the period of the QQO-like oscillation.}
\label{diffnf}
\end{figure*}

\begin{figure*}
\centering
\includegraphics[width=0.9\textwidth]{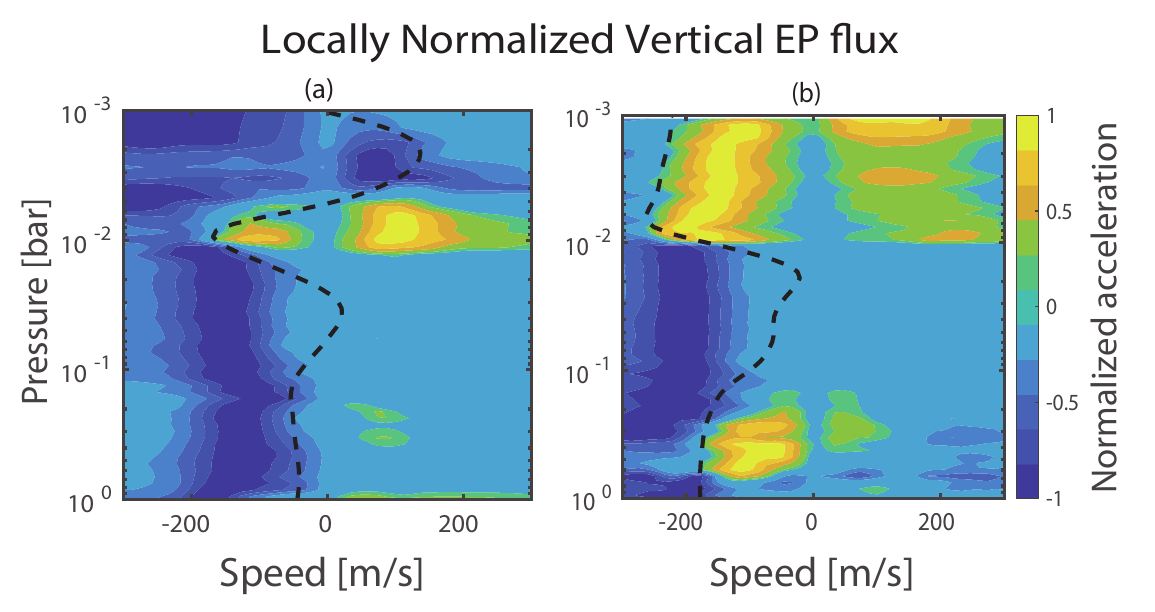}
\caption{Phase-speed spectra of the Eliassen-Palm flux at eastward wind phase (a) and westward wind phase (b) with $n_f = 10$ and $\tau_{\rm s} = 10^5$ ${\rm s}$, displaying the absorption of waves at critical levels. The equatorial zonal-mean zonal wind profiles at the corresponding times are overplotted in thick dashed curves. The spectra are normalized at every level by the maximum pressure at that level.}
\label{epflux}
\end{figure*}

\subsection{QQO-like oscillations at the Equatorial region}

In our simulations, we find the downward migrations of the stacked equatorial jets with alternating wind directions, exhibiting an oscillation similar to the QQO in Figure \ref{diffnf}, which shows the equatorial zonal winds with time for 10 Earth years over pressure-time planes. To keep the perturbations the same as that at the bottom, we confine the $f_{\rm amp}\sqrt{n_f}$ as a constant $1.3 \times 10^{-5}$ ${\rm K}$ ${\rm s^{-1}}$ (e.g. $f_{\rm amp} =$ $4 \times$ $10^{-6}$ ${\rm K}$ ${\rm s^{-1}}$ and $n_f =$ $10$). These cases are with $\tau_{\rm s}=10^5$ s and $\tau_{drag}=10^7$ s. The different cases exhibit complex structures with stacked jets of speeds from ${\rm -200}$ ${\rm m}$ ${\rm s^{-1}}$ to ${\rm 150}$ ${\rm m}$ ${\rm s^{-1}}$ between $10^{-3}$ bar and $0.1$ bar. The eastward phase tends to last longer in one oscillation period at larger pressure, but the westward phase is dominant at low pressure.

\begin{figure*}
    \centering
    \includegraphics[width=0.65\textwidth]{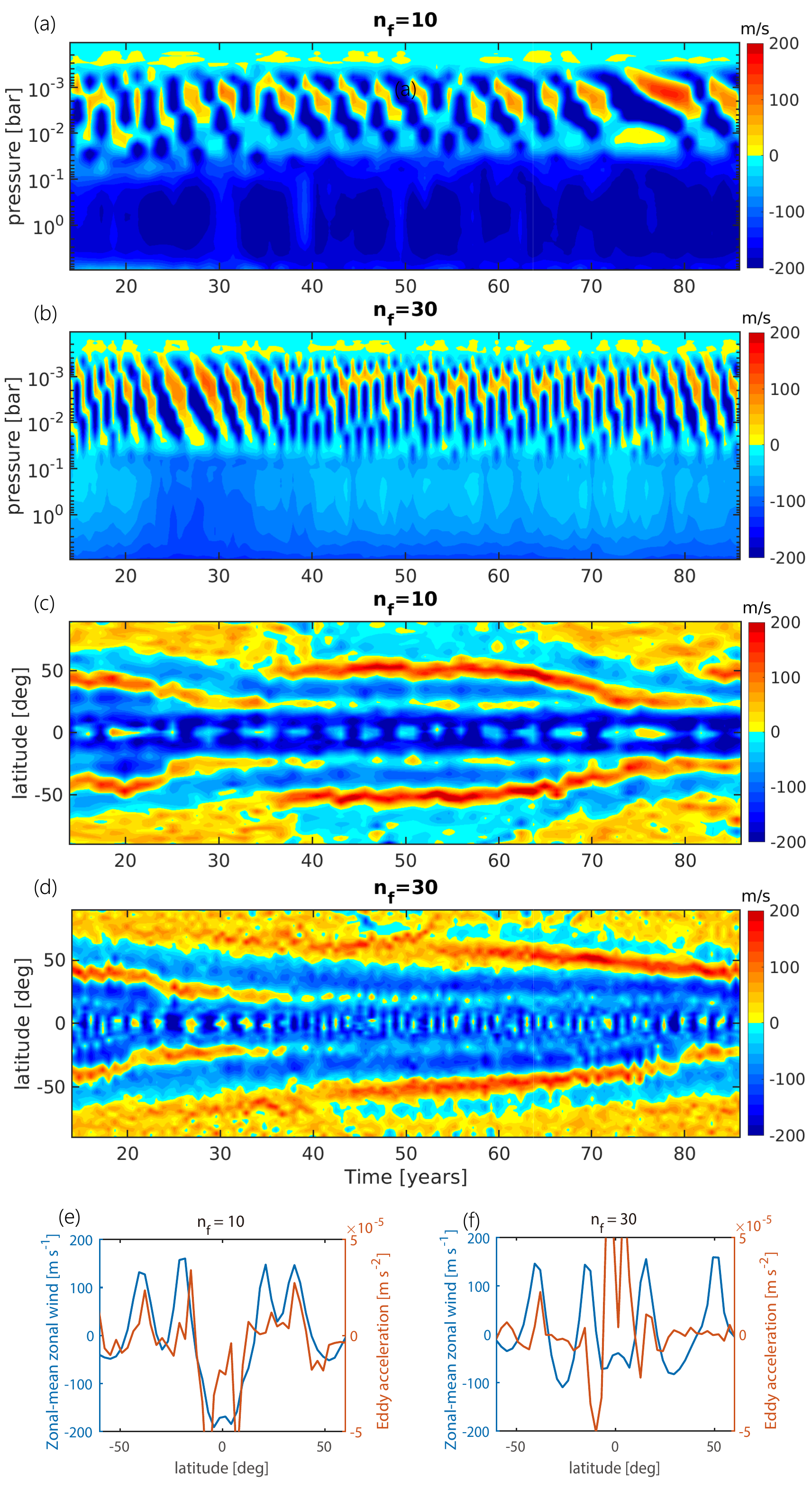}
    \caption{a,b: The equatorial zonal wind fields (time-pressure plane) for 90 Earth years with $n_f =$ 10, 30. c,d: The zonal wind fields (time-latitude plane) at 0.1 bar for 90 Earth years with $n_f =$ 10, 30. Red indicates eastward jets, and blue indicates westward jets. Color map units are m ${\rm s^{-1}}$. e,f: The zonal mean wind fields (blue line) and wave-induced accelerations (orange line) for the simulations run to 70 years with $n_f =$10, 30, smoothed by 20-day averaging. In all cases, the storm timescale $\tau_{\rm s} = 10^5$ s, and the drag timescale $\tau_{\rm drag} = 10^7$ s.}
    \label{qqo_long}
\end{figure*}

Different wavenumber cases have different periods of oscillation. Figure \ref{diffnf}b shows the oscillation periods with different forcing wavenumber $n_f$. The periods are defined as the time, which takes eastward wind changing to westward wind and then changing back to eastward wind at the level of $10^{-2}$ bar. Notice that the periods in Figure \ref{diffnf}b do not include the irregular periods, which are shown in Figure \ref{qqo_long}. 


Most QBO and QQO theories invoke that the equatorial waves accelerate the low wings of jets, causing the stacked jets to emerge over time and disappear at high pressure \citep{lindzen-etal-1968,holton-etal-1972,plumb-1977}. Here we apply diagnoses similar to \citet{showman-etal-2019} demonstrating that a similar mechanism is driving our simulated QQO-like oscillations.

\begin{figure*}
\centering
\includegraphics[width=1.0\textwidth]{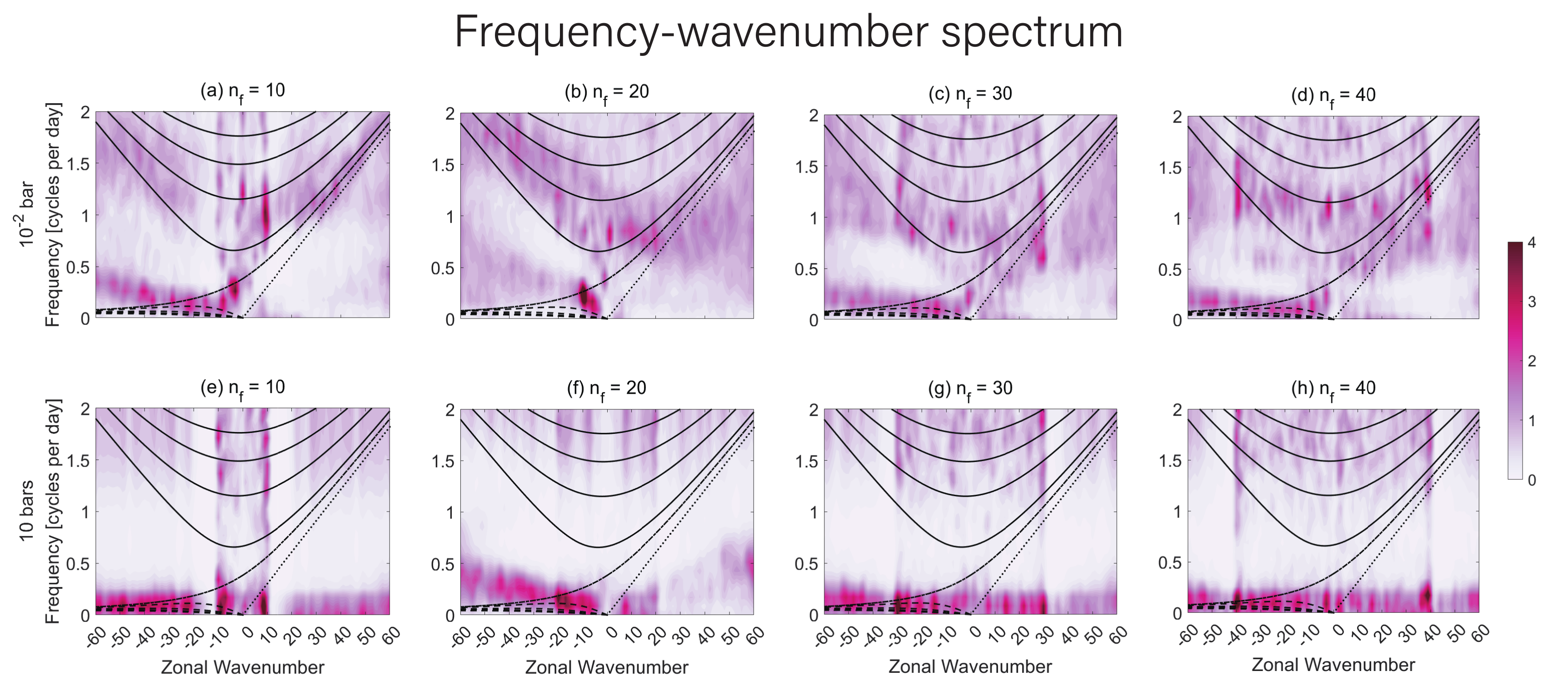}\\
\includegraphics[width=0.25\textwidth]{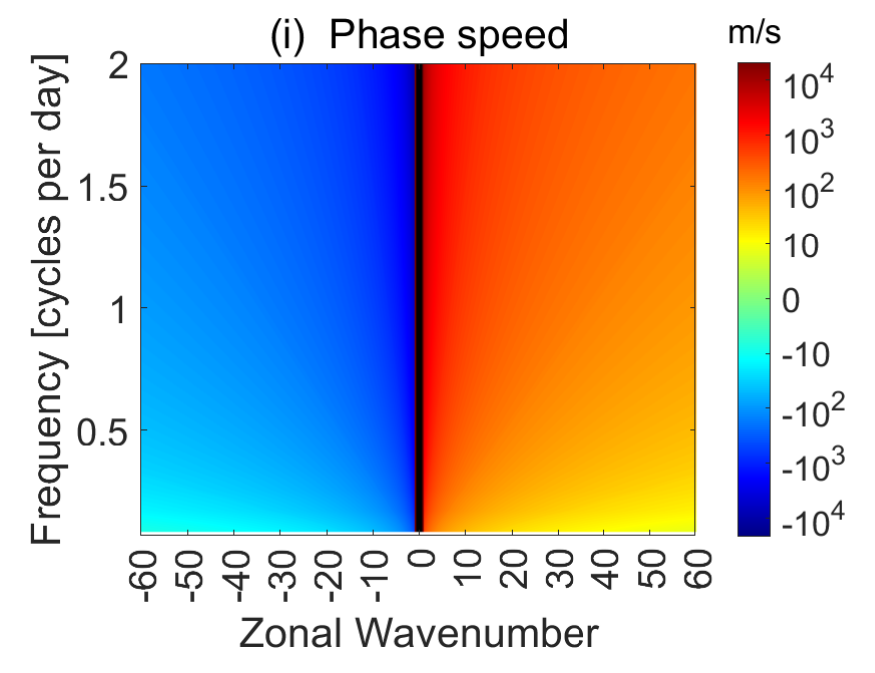}
\caption{Wavenumber-frequency spectra of temperature along equatorial region ($10^\circ {\rm S}-10^\circ {\rm N}$) at 10000 days exhibiting the QQO-like oscillations. X-axis gives a positive (negative) zonal wavenumber, representing eastward (westward) propagating waves, and Y-axis is the frequency in cycles per day (CPD). The colorscale indicates wave power density. Curves indicate analytic solutions of dispersion relations: Equatorial Rossby waves (dash lines), Mixed Rossby-gravity (MRG) waves (dash-dot lines), inertia-gravity waves (solid lines) and Kelvin waves (dot lines). The equivalent depths ($h_e$) of solutions are $1000$ m. Figure a-d show the Wheeler-Kiladis diagrams at $10^{-2}$ bar of different cases with forcing wavenumber from $10$ to $40$. Figure e-h show the Wheeler-Kiladis diagrams at $10$ bars of different forcing wavenumber cases. Figure i shows the phase speed by $\log_{10}$ scale.}
\label{fig:wk}
\end{figure*}

We recognize that the QQO-like oscillation is driven by the vertical momentum fluxes that are carried by the waves. The quasi-geostrophic approximation is no longer applicable at low latitudes, so the EP flux equation should be written in full form \citep{andrews-etal-1987}:
\begin{equation}
\begin{split}
\textbf{F}&={F}_{\phi}+{F}_z\\
&=\rho_0 R_J \cos{\phi}(\frac{\bar{u_z}\overline{v'\theta'}}{\bar{\theta}_z}-\overline{u'v'})\textbf{j}\\
&+\rho_0 R_J \cos{\phi}[(f-\frac{1}{a \cos{\phi}}\frac{\partial\bar{u}\cos{\phi}}{\partial\phi})\frac{\overline{v'\theta'}}{\bar{\theta}_z}-\overline{w'u'}]\textbf{k}   
\end{split}
\label{eqn:epfull}
\end{equation}
and the divergence of $\textbf{F}$ is:
\begin{equation}
\nabla \cdot \textbf{F}=\frac{1}{R_J\cos{\phi}}\frac{\partial(F_{\phi}\cos{\phi})}{\partial\phi}+\frac{\partial F_z}{\partial z}
\end{equation}

If the EP flux is divergent ($\nabla \cdot \textbf{F} > 0$), the mean flow obtains eastward momentum and has an eastward acceleration. Figure \ref{epflux} shows the vertical components of EP fluxes, which are normalized to the maximum absolute values at each pressure level, with the contours of zonal-mean winds. The EP fluxes are separated into the eastward modes (positive value) with speeds less than ${\rm 200}$ ${\rm m}$ ${\rm s^{-1}}$ and the westward modes (negative value) with speeds larger than ${\rm -200}$ ${\rm m}$ ${\rm s^{-1}}$. The wave absorption occurs at the critical levels near the bases of zonal jets, shown by the shading of EP fluxes ceasing extending when they encounter the dash lines, indicating the selective absorption effects of jets. The jets are favorable to absorb the waves that their phase speeds are close to the jet speeds. The absorption of eastward-propagating waves causes eastward accelerations and the absorption of westward-propagating waves causes westward accelerations, respectively. In the absence of damping, the waves with phase speeds far larger than the background flows could propagate upward to the top of the atmosphere without causing any alteration. We also notice an unexpected nature of the wave absorption remains in our analysis: some waves are also absorbed even if their speeds are faster than the background zonal-mean zonal winds. That phenomenon may be associated with the nonlinear effects, because the linear-wave-flow interaction theories generally assume small amplitudes of waves and may not completely capture all the details in our simulations.

Now we turn to characterize the wave properties. We apply a spectral analysis in the wavenumber-frequency domains about the equatorial waves \citep[e.g.][]{wheeler-etal-1999} in Figure \ref{fig:wk} of different forcing wavenumber cases at $10$ and $10^{-2}$ bar. We perform two-dimensional Fourier transforms on the temperature anomalies as a function of longitude and time to obtain the wavenumber-frequency characters at certain pressure levels. Then we obtain the final spectra from the coefficients divided by the background spectra, which are obtained from 40 times 1-2-1 filter smoothing of the raw coefficients. The frequency is cycles per day (CPD), which is the quotient of the angular frequency and $2 \pi$. The positive zonal wavenumber means the eastward propagation of waves and the negative zonal wavenumber means westward propagation.

We would like to lay out the theories of equatorial free waves for comparisons of wave properties. The analytic solutions help us classify the waves in Figure \ref{fig:wk}. Considering a linearized shallow water equation for perturbations on an equatorial $\beta$-plane that $f \approx \beta y$, the analytic solutions of equatorial waves in one layer are written as Equation \ref{eqn:wave} \citep{matsuno-1966}: 

\begin{equation}
\frac{c_K}{\beta}(\frac{-k\beta}{x}-k^2+\frac{x^2}{c_K^2})=2n+1
\label{eqn:wave}
\end{equation}

The curves show dispersion relationships of different equatorial waves, where $c_K$ is the Kelvin wave phase speed, $k$ is the zonal wavenumber, and $n = 0, 1, 2, 3...$. When $n = 0$, we obtain the solutions of the mixed Rossby-gravity (MRG) waves (dash-dot lines); When $n = 1, 2, 3...$, we obtain the solutions of the inertia-gravity waves (solid lines) and the Rossby waves (dash lines); Dot lines show the equatorial Kelvin waves with wavenumber-frequency relationship $\omega = kc$ with equivalent depths $h_e = 1000$ m. The equivalent depths are theoretical layer depths used in shallow-water models to specify an intrinsic wave speed in the shallow water system such that $c_K=\sqrt{gh_e}$. We tune the $h_e$ to make the theoretical curves closer to our calculated spectral signals in Figure \ref{fig:wk}.

We compare each figure in Figure \ref{fig:wk}: Rossby wave modes (dash lines) lie in low-frequency regions at $10$ bars, with phase speeds ranging from $-5$ ${\rm m}$ ${\rm s^{-1}}$ to $-40$ ${\rm m}$ ${\rm s^{-1}}$, indicating that the Rossby wave modes could be directly generated by the thermal perturbations in our simulations. The Rossby wave modes increase their phase speed in the $10^{-2}$ bar panel, showing the loss of energy at the critical levels when low-frequency-Rossby waves propagate upward through the equatorial jets.

The Kelvin wave modes are shown by dot-dash lines. In the lower panel of Figure \ref{fig:wk}, there are only slow Kelvin waves with phase velocities of more than 10 m ${\rm s^{-1}}$; In the upper panel of Figure \ref{fig:wk}, the slow Kelvin waves disappear. Most Kelvin modes are with phase velocities $\sim$ 150 m ${\rm s^{-1}}$, which also illustrates that the slow Kelvin waves are filtered by the background flow.

Inertia-gravity waves (solid lines) are present in all cases, indicating nonlinear effects in our experiments: in the lower panel of Figure \ref{fig:wk} with 10 bar, the inertia-gravity wave power concentrates around the forcing wavenumber $n_f$ , and in the upper panel of Figure \ref{fig:wk}, the inertia-gravity wave power disperse to wider wavenumbers and frequency regions, indicating that the energy is transferred from the forcing wavenumber $n_f$ to the other wavenumbers. The inertia-gravity wave speed is around 200 to 400 m ${\rm s^{-1}}$, which is greater than the wind speed around 150 m ${\rm s^{-1}}$ in Figure \ref{fig:sph-drag}, indicating that most of the inertia-gravity waves with phase velocities smaller than the background flow have been absorbed below the critical levels, while the faster inertia-gravity waves are transparent to the background flow, so they can propagate to regions with lower pressure.

The MRG waves exhibit the strongest amplitudes in the top panel in all cases, especially the westward propagating waves with zonal wavenumbers between $-5$ and $-11$. Their phase speeds are in range from $-40$ to $-400$ ${\rm m}$ ${\rm s^{-1}}$. These MRG waves are easily penetrating the background flow and reach the top of the atmosphere, and they may resemble the equator-trapped planetary-scale waves that have been observed in \citet{allison-1990,deming-etal-1997}. In particular, the power of the MRG waves with $n_f =$ 10 and 20 is larger than the wave power with $n_f =$ 30 and 40.

Based on the above wave analysis, we suggest a hypothesis to explain the shortening of QQO-like oscillation as a function of forcing wavenumber $n_f$ in Figure \ref{diffnf}. The jet center of the QQO-like oscillation is located at $10^{-2}$ bar, so the more waves are absorbed at this position, the wave-induced accelerations are greater, and the periods of the QQO-like oscillation are shorter. Essentially, most upward-propagating waves are generated by the cascade of the waves, which are excited by internal forcing, and their wavenumbers are related to the forcing wavenumber $n_f$: larger $n_f$ corresponds to the smaller wavelength, exciting more small-scale gravity waves. These inertia-gravity waves may contribute significantly to the QQO momentum budgets, as shown in Figure \ref{fig:wk}a, b, c, and d. Compared with each other, the distributions of inertia-gravity waves in Figure \ref{fig:wk}d are the widest in ranges and correspond to the darkest color, while the inertia-gravity waves in Figure \ref{fig:wk}a are obviously less widely distributed than which in d. Therefore, a larger $n_f$ excites more small-scale inertia-gravity waves, and the small-scale waves are absorbed, resulting in an increase in the momentum acquired by the background flow, and thus the QQO-like oscillation period becomes shorter.

In particular, we notice variable periods of the QQO-like oscillations when the simulations continue running. Figures \ref{qqo_long}a, b show the evolutions of QQO-like oscillations lasting 90 years with $n_f =$ 10 and 30. It can be seen that the deep equatorial westward jets at 0.1 bar are gradually intensified in Figures \ref{qqo_long}a. After the simulations reach about 80 years, the maximum deep equatorial westward jet speeds are -200 m ${\rm s^{-1}}$. At this time, the average periods are extended from $\sim$ 3.4 years to $\sim$ 5 years. The same results occur in Figure \ref{qqo_long}b: when the deep equatorial westward jet speeds are maintained at about -40 m ${\rm s^{-1}}$, the periods of the QQO-like oscillations last less than 2 years, while the simulations run between 20 and 40 years, the deep equatorial westward jet speeds increase to $\sim$ -100 m ${\rm s^{-1}}$, correspondingly, the periods of the QQO-like oscillations also extend to about 3 years.

As can be seen in Figure \ref{qqo_long}c, d, the evolution of the deep equatorial jets and the off-equatorial jets are directly related. After the simulations run for 20 to 30 years, and after 80 years, the off-equatorial jets migrate to equatorial regions, and the momentum exchanges strengthen the deep equatorial westward jets, which filter the upward-propagating equatorial waves, and lead to longer periods of QQO-like oscillations. These results indicate that the change of periods of the QQO-like oscillations in our simulations is related to the interactions between the off-equatorial migrating jets and the equatorial jets.



The migrations of the off-equatorial jets have also been mentioned in some idealized GCM studies \citep[e.g.][]{feldstein-1998,robinson-2000,chan-etal-2007,chemke-kaspi-2015,ashkenazy-tziperman-2016}. These migrations are believed caused by that the centers of eddy-induced wave accelerations and the jet centers do not coincide, which are similar to the Figure \ref{qqo_long}e,f in our simulations. The physical scenario of the variation of QQO's period in our simulations is now clear: the equatorward acceleration biases on the off-equatorial jets cause equatorward migrations, and these jets interact with the deep equatorial westward jets and enhance them; The deep jets filter the upward propagating equatorial waves, reducing the EP fluxes in the oscillation layers and prolonging the oscillation periods. However, we are cautious that the latitudinal jet migrations that occurred in idealized GCMs  have not been observed in real atmospheres. The reason may include the simplified parameterizations of the forcing and the lack of some crucial physical processes in idealized models, such as the feedback on the forcing patterns by the emergence of strong jets, water vapor, and moist convection.

\subsection{Sensitivity tests of the QQO-like oscillations}

\begin{figure*}
\centering
\includegraphics[width=0.9\textwidth]{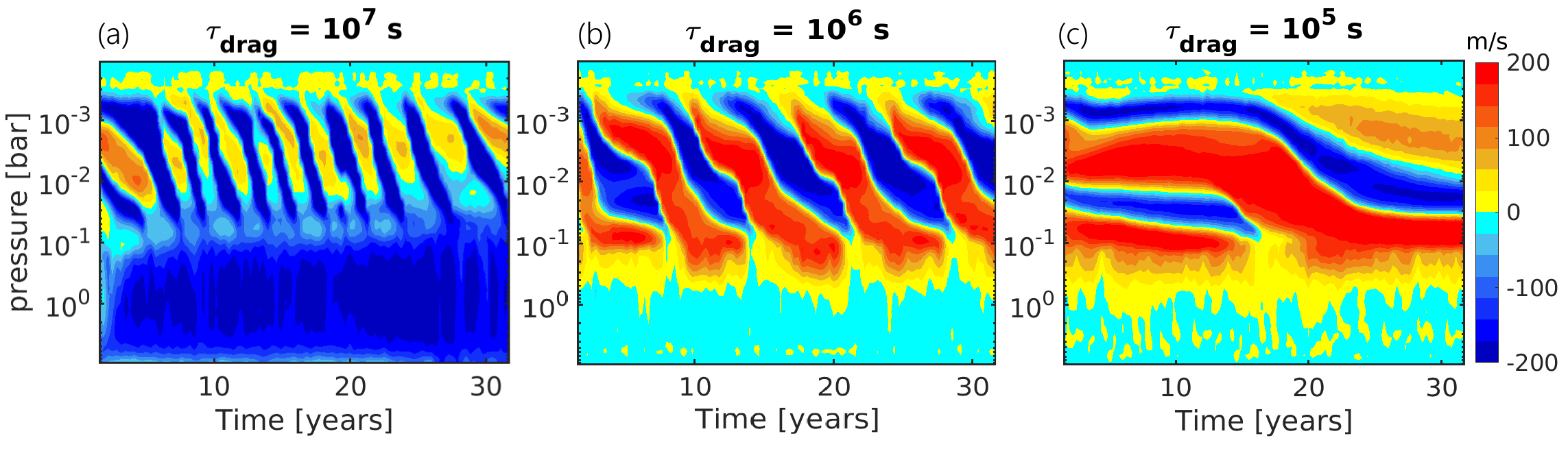}
\caption{Time varying equatorial zonal winds, with different drag timescales. Red color denotes eastward zonal winds, and blue color denotes westward zonal winds. In all simulations, the forcing amplitude is $f_{\rm amp} = 4 \times 10^{-6}$ ${\rm K}$ ${\rm s^{-1}}$, the zonal forcing wavenumber is $n_f = 10$, and the deacy timescale is $\tau_{\rm s} = 10^5$ ${\rm s}$. Color bar unit is ${\rm m}$ ${\rm s^{-1}}$.}
\label{fig:diffdrag}
\end{figure*}

\begin{figure*}
\centering
\includegraphics[width=0.9\textwidth]{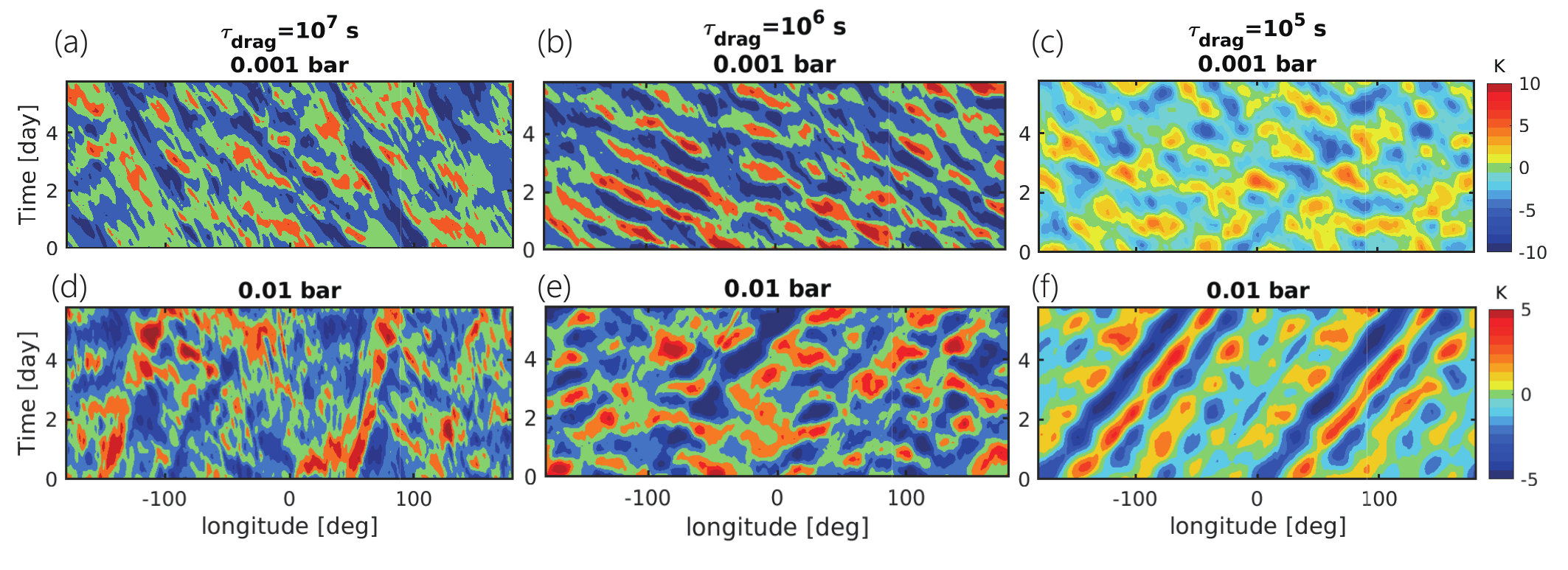}
\caption{Time-longitude cross sections of equatorial temperature anomalies at equilibrium states. Top panels: 0.001 bar, and bottom panels: 0.01 bar. The parameters are same as that in Figure \ref{fig:diffdrag}.}
\label{fig:wavedrag}
\end{figure*}

In view of our limited computing resources, we tested the case $n_f =10$, which has the oscillation periods closer to Jupiter’s QQO period, with the different bottom drag ($\tau_{\rm drag}$), time correlation ($\tau_{\rm s}$) and initial zonal winds.

\subsubsection{bottom drag}

\begin{figure*}
\centering
\includegraphics[width=0.95\textwidth]{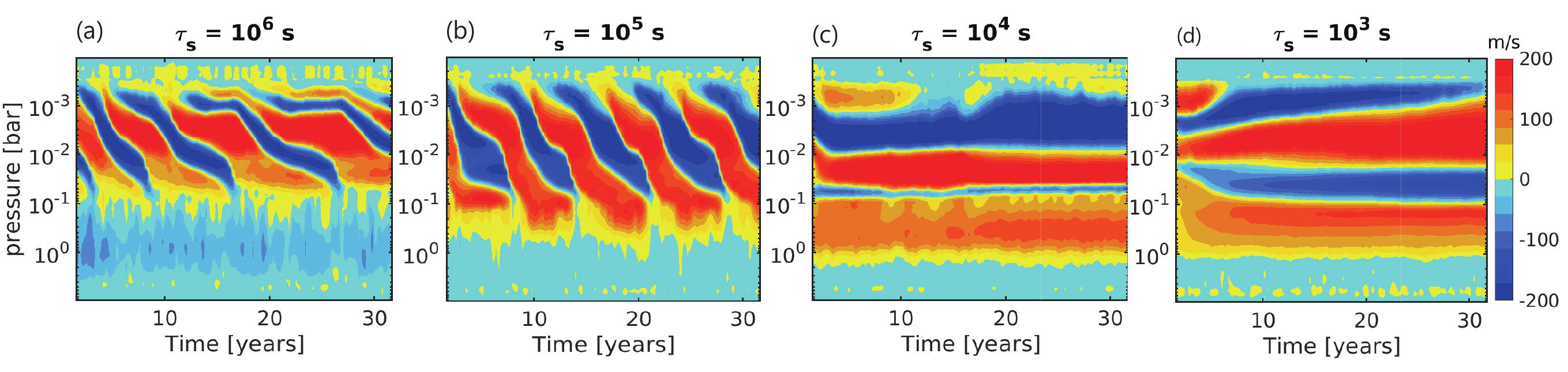}
\caption{Time varying equatorial zonal winds, with different storm timescales. Red denotes eastward winds, and blue denotes westward winds. In all simulations, the forcing amplitude is $f_{\rm amp} = 4 \times 10^{-6}$ ${\rm K}$ ${\rm s^{-1}}$, the zonal forcing wavenumber is $n_f = 10$, and drag timescale is $\tau_{\rm drag} = 10^6$ ${\rm s}$. Color bar unit is ${\rm m}$ ${\rm s^{-1}}$.}
\label{fig:difftaud}
\end{figure*}

The development of prograde (eastward) and retrograde (westward) deep equatorial jets likely have significant influences on the wave properties and the QQO-like oscillations, causing the lengthening of periods (like Figure \ref{diffnf}a, $n_f =20, 25$ case). That could be caused by the filtering effects of the equatorial jets on the wave properties. We increase the bottom drag to suppress the deep equatorial jets.

Figure \ref{fig:diffdrag} shows the equatorial zonal winds as a cross-section of time and pressure. The stronger bottom drag ($\tau_{\rm drag}$ decrease from $10^7$ ${\rm s}$ to $10^5$ ${\rm s}$), the longer period of the QQO-like oscillation (from about $3$ Earth years to about $20$ Earth years), the phenomenon illustrated the bottom drag may restrict the equatorial waves and weaken the momentum flux. The equatorial jet at $1$ bar disappears with a strong bottom drag, but the wind speed of the QQO-like oscillation does not change substantially. Then we show the temperature anomalies in Figure \ref{fig:wavedrag}. Comparisons between the panels of Figure \ref{fig:wavedrag} show that the equatorial waves are weaker with increasing bottom drag. The wave crests separate in large-drag simulations, illustrating that the waves lost high-frequency features and large-wavenumber features when the bottom drag is strong. 

The disappearance of the small-scale waves could be an explanation for the longer period oscillation in Figure \ref{fig:diffdrag}. Both cases with $\tau_{drag}=10^6$ and $10^5$ s have weak deep jets, and the QQO-like oscillation period is quite different. Although strong bottom drag suppresses the deep equatorial jets, causing the equatorial waves to propagate much more transparently out of the bottom layers, it damps the velocity anomalies, either. The strong drag in the case with $\tau_{drag}=10^5$ also damp out some waves generated in the deep layers, and the EP fluxes are relatively weak to accelerate the stratosphere oscillation. Finally, stronger drag leads to a longer QQO-like oscillation period.

\subsubsection{time correlation}
\label{time}

We then conduct simulations with different storm timescales $\tau_{\rm s}$ from $10^6$ s to $10^3$ s. $\tau_{\rm s}$ shows the injection frequency of bottom perturbation, characterizing the wave properties in the time axis. Figure \ref{fig:difftaud} shows the equatorial zonal winds as a function of time, and a bottom drag timescale $\tau_{\rm drag}=10^6$ s is applied to prevent the formation of strong deep equatorial jets. Figure \ref{fig:difftaud} implies that the deep zonal jets develop when the storm timescale decreases. As eastward jets develop, the QQO-like oscillations signature disappears. 

\begin{figure}
    \centering
    \includegraphics[width=0.4\textwidth]{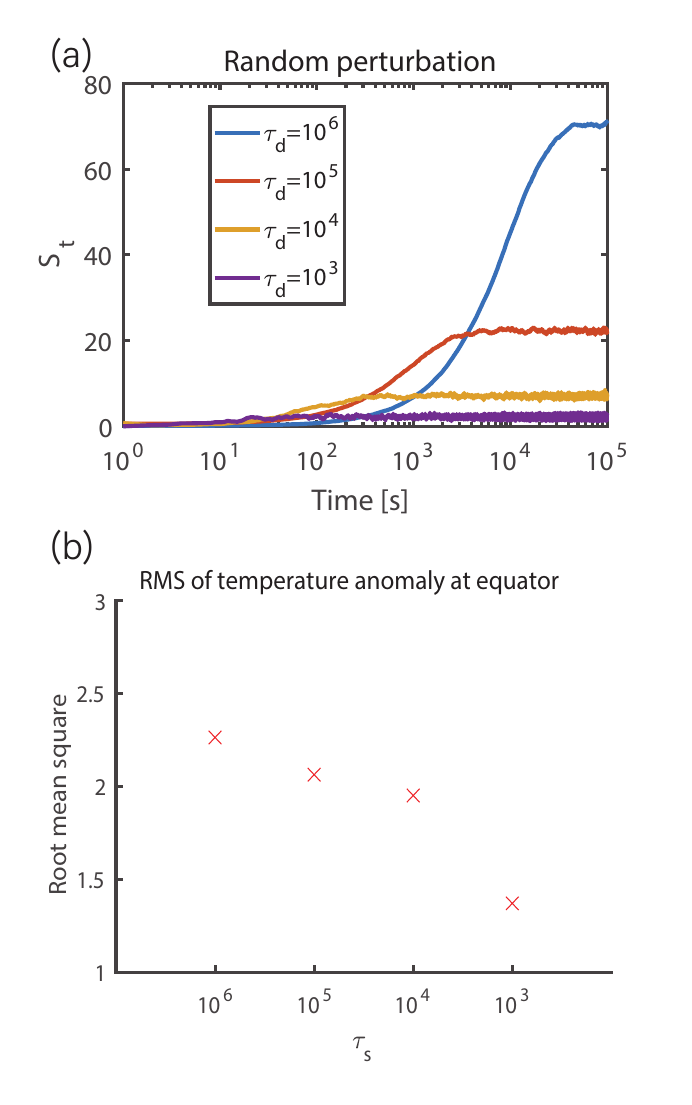}
    \caption{Influences of $\tau_{\rm s}$ on perturbations and waves. (a) Time-varying local perturbations with different storm timescale}s, and (b) root mean square of equatorial temperature anomalies ($10^\circ$ S-$10^\circ$ N).
    \label{fig:perturbation}
\end{figure}

We speculate that the varying $\tau_{\rm s}$ could have an influence on the wave flux. Figure \ref{fig:perturbation}a shows the sum of the heating rate, providing a view of time-depend perturbations. We focus on one air parcel overlying the deep layer, which is heated by the interior convection, and the ``accumulate heating rate'' is $S_t$, parameterized by Equation \ref{pertur}:
\begin{equation}
    S_t(t+\delta t)=\sqrt{1-\alpha^2}S_t(t)+\alpha x_m
\label{pertur}
\end{equation}
$\alpha$ is de-correlation factor in Equation \ref{markov} depend on time step $\delta t$ and $\tau_{\rm s}$. $x_m$ is a random number in the range of [-1,1], that the negative value represents cooling, and the positive value represents heating. The result in Figure \ref{fig:perturbation}a implies that, without considering radiative cooling, the larger $\tau_{\rm s}$, the stronger parameterized convective heating. This result is further illustrated by Figure \ref{fig:perturbation}b: The root mean square (RMS) of temperature anomaly at equator regions decreases with a decline of $\tau_{\rm s}$ at equatorial region, suggesting less equatorial wave activities.

The RMS of temperature anomalies helps to explain the disappearances of the QQO-like oscillations of short $\tau_{\rm s}$ case in Figure \ref{fig:difftaud}. The wave power and the temperature anomalies are positively correlated. We could expect that the wave power to be weaker when the $\tau_{\rm s}$ is shorter. The weaker waves are weakened by absorption through the jets and the radiative damping. By the time they reach the top, there is not enough wave power left to generate a new stacked jet. A result is that the jet structures keep steady over time when $\tau_{\rm s}$ is small.

\subsubsection{initial zonal winds}
\begin{figure}
    \centering
    \includegraphics[width=0.4\textwidth]{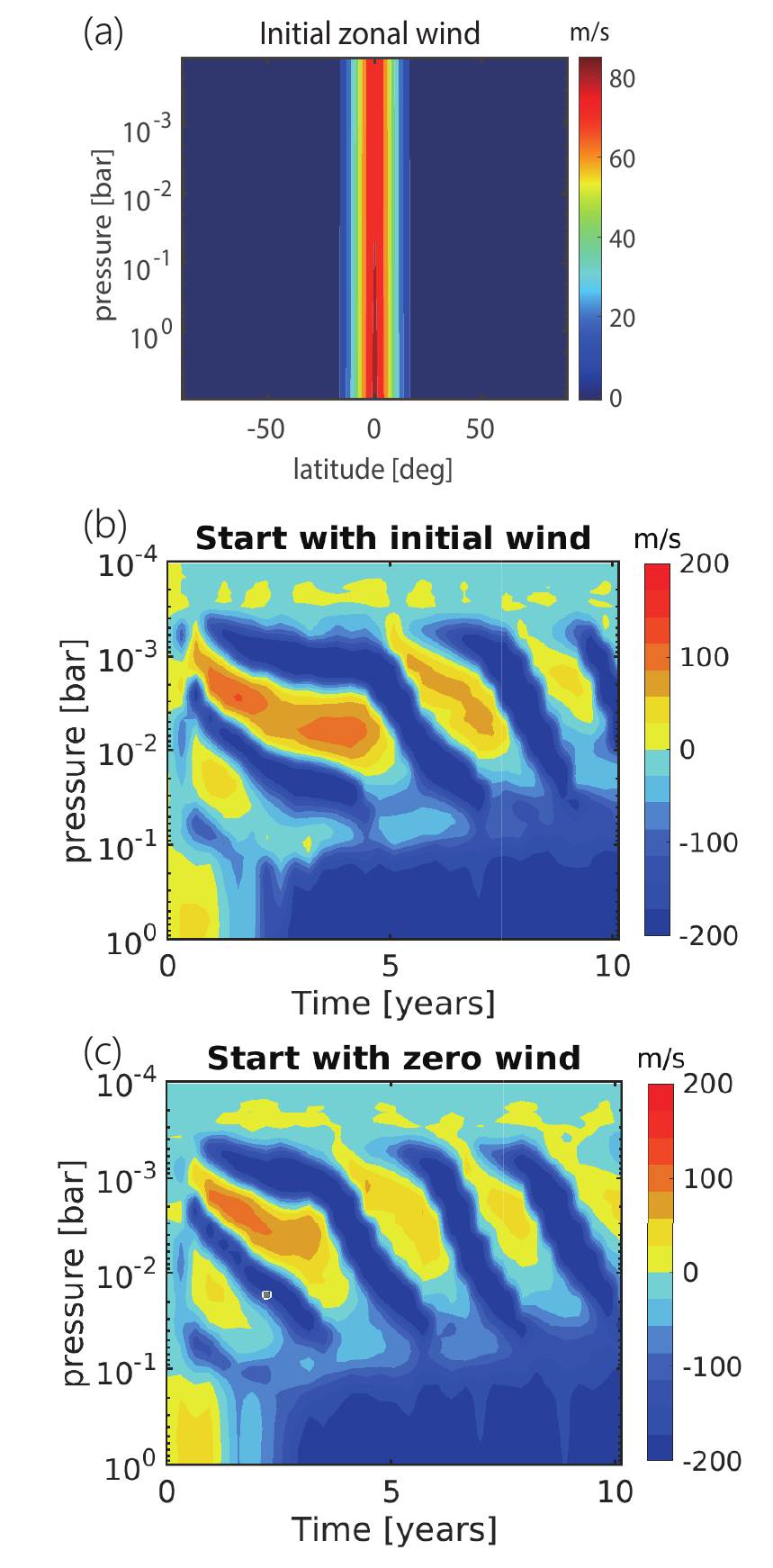}
    \caption{(a) Vertical structures of initial zonal wind profiles, simplified from the Hubble Space Telescope observations \citep{simon-etal-2015}. Note that tropical zonal wind speeds decrease with height. (b) The zonal winds at the equator start with the initial zonal wind of a. (c) The zonal winds at the equator start with zero wind.}
    \label{fig:init}
\end{figure}

We initialized our $n_f = 10$ case with zonal eastward wind from the observations in Figure \ref{fig:init}a. The initial wind profile is a Gaussian equatorial jet, with a maximum speed of $81$ ${\rm m}$ ${\rm s^{-1}}$ and the full width at half maximum of $\pm 8^\circ$.

Figure \ref{fig:init}b and c show the QQO-like oscillation behavior with initial eastward zonal wind and zero wind at the beginning, respectively. Results show that the wind shears descend slower with initial eastward winds. Except for a $\sim 2$ year lag in Figure \ref{fig:init}b compared to Figure \ref{fig:init}c, the QQO-like oscillation periods are the same in Figure \ref{fig:init}b and c, which are about 3.5 Earth years.

These results suggest that the final states are independent of the initial conditions. The initial wind structures do not participate in the wave-flow interaction, except for some period lags. However, the formation mechanism of Jupiter's deep eastward equatorial jets is still an open question and is out of the scope of our current study.

\section{Conclusions and Discussion}
\label{sec:conclusions}

We have presented idealized 3D general circulation models of Jupiter to investigate the formation of zonal jets and the evolution of QQO-like oscillations. We have also performed sensitivity tests to help understand the dynamical origins of the jets and simulated QQO-like oscillations. We have the primary results as follows:

$\bullet$ Our simulations showed that zonal jets emerge naturally in Jupiter's conditions with isotropic internal forcing. The jets are self-organized with speeds in a range of $\sim \pm 100$ ${\rm m}$ ${\rm s^{-1}}$. The mechanism of jet formation is related to the meridional propagation of Rossby waves and the inhomogeneous potential vorticity evolution. The wind speed of the jets, as well as the number of the jets, decrease with increasing bottom drag strength. The off-equatorial jets are more sensitive to drag and radiative damping than the equatorial jets.

$\bullet$ In our present work, the 3D simulations generate QQO-like oscillations in the equatorial stratosphere, which are believed to be driven by the interaction of upward-propagating waves from the troposphere into the stably stratified stratosphere. The stacked eastward and westward jets migrate downward, and the periods of oscillation shorten with the increasing forcing wavenumber. We show that the dominant forcing wave modes with large-scale forcing wavenumber $n_f = 10$ produce the closest QQO-like oscillation behavior to the observed QQO periodicities. Besides, the mixed Rossby-gravity waves with a wavenumber of $10$ may have a large possibility of influencing the upper layer, exhibiting features similar to observations of cloud plumes.

$\bullet$ We show that the deep equatorial jets have strong influences on the QQO-like oscillations. The migrations of off-equatorial jets increase the strength of the deep equatorial jets, resulting in the prolonging of the QQO-like oscillations shown in our models. Our simulations are independent of the initial conditions. Our simulations are also sensitive to the bottom drag. Detailed diagnostics show that the bottom drag inhibits the formation of deep jets, and it also damps the equatorial waves at the deep layers where they are generated, resulting in longer QQO-like oscillations. The parameters storm timescale $\tau_{\rm s}$ are associated with the wave power. The shorter storm timescale $\tau_{\rm s}$, the smaller wave power, result in weakening and disappearance of the off-equatorial jets.

Our results support that the upward propagating equatorial waves trigger the QQO-like oscillation as suggested by classical theories \citep{lindzen-1970,lindzen-1971,lindzen-1972}. Our model shows that the simulation periods close to Jupiter's QQO, are most likely to be generated by large-scale wave forcing with $n_f=10$, qualitatively in agreement with some simulations. \citet{allison-1990} shows that the Rossby modes exhibit the largest amplified rate around zonal wavenumber $10$. \citet{li-1993,li-read-2000,li-etal-2006} also exhibit the scale selections of the dominant Rossby wave modes with wavenumber $1 \pm 15$. And the large-scale waves are also consistent with the long-term infrared hot spot observations \citep[e.g.][]{hunt-etal-1981,ortiz-etal-1998,choi-etal-2013}. 

We can compare qualitatively our results to the previous work and the observations. \citet{friedson-1999} estimate a descent rate of QQO wind shear of $0.16$ ${\rm cm}$ ${\rm s^{-1}}$. \citet{cosentino-etal-2017} determined the descent rate of $\sim 0.04$ ${\rm cm}$ ${\rm s^{-1}}$ in their model and $\sim 0.05$ ${\rm cm}$ ${\rm s^{-1}}$ from TEXES data. Our best-match model shows the descent rate of $0.039$ ${\rm cm}$ ${\rm s^{-1}}$ to $0.063$ ${\rm cm}$ ${\rm s^{-1}}$. The descent rate is related to the EP flux carried by the waves. \citet{friedson-1999} overestimated the EP fluxes of $\sim 7 \times$ $10^{-4}$ ${\rm m^2}$ ${\rm s^{-2}}$, while they span a range from $\sim 1 \times$ $10^{-4}$ to $\sim 4 \times$ $10^{-4}$ ${\rm m^2}$ ${\rm s^{-2}}$ in our simulations. 

Our models show Rossby wave modes with phase speeds of $-10 \sim -50$ ${\rm m}$ ${\rm s^{-1}}$ and MRG wave modes with phase speeds of $-40 \sim -400$ ${\rm m}$ ${\rm s^{-1}}$.
Although the large-scale waves contribute mostly to generating the QQO, small-scale gravity waves that cannot be resolved in global models could contribute to shaping the QQO's structures and periods \citep{cosentino-etal-2017}, and they are in general essential to force the QBO-like oscillations \citep{baldwin-etal-2001}. Planetary waves occur over a broader range of latitudes off equator \citep{orton-etal-1994,deming-etal-1997}, but the inertia-gravity waves may contribute a significant part to the QQO momentum budgets \citep{cosentino-etal-2016,cosentino-etal-2017b}, due to they are no equatorial confinement. Large-scale equatorial waves could be trapped in narrow equatorial waveguides, while small-scale gravity waves don't suffer this restraint \citep{friedson-1999}. Kelvin and MRG waves are equatorially restrained within $\pm 5^\circ$ in our simulations, however, the equatorial jets extend to about $\pm 15^\circ$. If much higher horizontal resolutions are applied, the simulations may resolve more small-scale gravity waves, strengthening the QQO signals as they did in the QBO models \citep[e.g.][]{hayashi-golder-1994,takahashi-1996}. Increasing higher vertical resolutions also leads to larger amplitude of the MRG, the Kelvin waves and the EP fluxes \citep{richter-etal-2014}. 

The behaviors of sub-grid gravity waves that cannot be explicitly captured in GCMs are complex, and the wave-breaking process is affected by many aspects, including the background wind shears \citep{hines-1991}, the diffusive timescales \citep{lindzen-1981}, the onsets of convective instability \citep{smith-etal-1987} and the mixing process of air parcels that do not return the original position \citep{medvedev-klassen-1995}. It is difficult to characterize these sub-grid behaviors quantitatively in our simulations, which could influence the gravity waves and the QQO-like oscillations, a task we leave to the future.

The common outcome of westward equatorial jets in the troposphere seems to be a general property of the stochastic, isotropic forcing, which occurs in many previous shallow-water models \citep{scott2007forced,showman2007numerical, zhang2014atmospheric} and primitive equation models \citep{showman-etal-2019,tan-2022}, including models in this study. This is in stark contrast to the strong and stable equatorial superrotation observed in Jupiter's and Saturn's tropospheres. This suggests that some fundamental feedback mechanisms may be missing in the stochastic forcing framework. On the other hand, major mechanisms controlling the equatorial superrotation and general circulation in the tropospheres of Jupiter and Saturn are unsettled and still under active investigation (see a recent review of \citealp{showman2018global}). Generating a self-consistent equatorial superrotation like the observed ones on Jupiter and Saturn is an important issue in the field of planetary atmosphere and is  out of our current scope. We expect that if the tropospheric equatorial superrotation is self-consistently maintained in our models, it will have a quantitative influence on the simulated QQO because of the wave-filtering effects of the jet.  Nevertheless, given that our current study has focused on a mechanism study of QQO, the wave-mean-flow interaction mechanism discussed here should still hold.

Jupiter's internal heat flux and the interactions with the overlaying stratified layers may not be strictly isotropic and may exhibit certain latitudinal variations. In addition, the QQO is likely to be disrupted by planetary-scale disturbances in the equatorial and low-latitude troposphere \citep{antunano-etal-2021}. These warrant future numerical explorations of how latitudinal-dependent internal forcing would influence the general circulation and the QQO of Jupiter.   This work has focused on the isotropic  internal forcing  for a few reasons. First, the internal heat flux of Jupiter is expected to be horizontally homogeneous on the zeroth order \citep{ingersoll1978solar,ingersoll2004dynamics,fortney-nettelmann-2009}. Second,   the stochastic and isotropic nature of the forcing allows us to understand the formation and time evolution of zonal jets in a clean setup by excluding latitudinal forcing variations. This ties our work to the rich literature investigating turbulence and jet formation in the context of  giant planets' atmospheres using isotropic turbulent forcing (see reviews of \citealp{vasavada2005jovian,galperin2008zonostrophic,showman2018global} and many references therein). Third, there are several candidate mechanisms responsible for latitudinal-dependent internal forcing, each could imply different latitudinal forms of the forcing. For instance, rapidly-rotating-shell convection models (e.g., \citealp{showman-etal-2013}) suggest that convection tends to be organized into columns parallel to the rotation axis and  convective velocities are expected to be greater at high latitudes; even if convective heat flux is isotropic, the perturbations caused in the stratified layers may be stronger at low latitudes where the horizontal divergence of the large-scale motions is on a leading order \citep{schneider2009formation}; if moist convection is the leading mechanism of internal heat transport in the troposphere of Jupiter \citep{gierasch2000observation, ingersoll-2000}, storms associated with moist convection and latent heating could be enhanced at low latitudes and be self-organized and coupled with the large-scale flows \citep{lian-showman-2010}. That said, given the multiple origins of the latitudinal internal forcing variations, a variety of latitudinal forcing forms should be investigated in future studies to understand the sensitivity of circulation and QQO. Therefore,  our study with an isotropic forcing serves as a baseline study for comparisons with future Jovian atmospheric models that incorporate (the uncertain) latitudinal forcing variation.

\section*{Acknowledgements}
This work is supported by the National Natural Science Foundation of China, under Grants No. 41888101. Numerical simulations were conducted at the High-performance Computing Platform of Peking University.

\bibliography{ref}{}
\bibliographystyle{aasjournal}

\end{document}